\documentclass{article}

\usepackage{microtype}
\usepackage{graphicx}
\usepackage{subfigure}
\usepackage{booktabs} 

\usepackage{hyperref}

\usepackage[accepted]{icml2023}

\usepackage{amsmath}
\usepackage{amssymb}
\usepackage{mathtools}
\usepackage{amsthm}

\usepackage[capitalize,noabbrev]{cleveref}

\theoremstyle{plain}

\theoremstyle{definition}

\theoremstyle{remark}

\usepackage[textsize=tiny]{todonotes}

\usepackage{amsmath}
\usepackage{amssymb}
\usepackage{mathtools}
\usepackage{amsthm}
\usepackage{multirow}
\usepackage{tabularx}
\usepackage[capitalize,noabbrev]{cleveref}
\usepackage{algorithmic}

\usepackage[textsize=tiny]{todonotes}

\icmltitlerunning{Attention-Based Recurrence for Multi-Agent Reinforcement Learning under Stochastic Partial Observability}

\begin{document}

\twocolumn[
\icmltitle{Attention-Based Recurrence for Multi-Agent Reinforcement Learning under Stochastic Partial Observability}



\icmlsetsymbol{equal}{*}
\icmlsetsymbol{worklmu}{\textdagger}

\begin{icmlauthorlist}
\icmlauthor{Thomy Phan}{usc,worklmu}
\icmlauthor{Fabian Ritz}{lmu}
\icmlauthor{Philipp Altmann}{lmu}
\icmlauthor{Maximilian Zorn}{lmu}
\icmlauthor{Jonas Nüßlein}{lmu}
\icmlauthor{Michael Kölle}{lmu}
\icmlauthor{Thomas Gabor}{lmu}
\icmlauthor{Claudia Linnhoff-Popien}{lmu}
\end{icmlauthorlist}

\icmlaffiliation{usc}{University of Southern California,  USA.  \textsuperscript{\textdagger}Work done at LMU Munich}
\icmlaffiliation{lmu}{LMU Munich,  Germany.  This paper is an extension of \cite{phan2022attention2}}
\icmlaffiliation{lmu}{LMU Munich,  Germany.  This paper is an extension of \cite{phan2022attention2}}

\icmlcorrespondingauthor{Thomy Phan}{thomy.phan@ifi.lmu.de}

\icmlkeywords{Dec-POMDP, Stochastic Partial Observability, Multi-Agent Learning, Recurrence, Self-Attention}

\vskip 0.3in
]



\printAffiliationsAndNotice{}  

\begin{abstract}
Stochastic partial observability poses a major challenge for decentralized coordination in multi-agent reinforcement learning but is largely neglected in state-of-the-art research due to a strong focus on state-based \emph{centralized training for decentralized execution (CTDE)} and benchmarks that lack sufficient stochasticity like \emph{StarCraft Multi-Agent Challenge (SMAC)}.
In this paper,  we propose \emph{Attention-based Embeddings of Recurrence In multi-Agent Learning (AERIAL)} to approximate value functions under stochastic partial observability.  AERIAL replaces the true state with a learned representation of multi-agent recurrence,  considering more accurate information about decentralized agent decisions than state-based CTDE.
We then introduce \emph{MessySMAC},  a modified version of SMAC with stochastic observations and higher variance in initial states,  to provide a more general and configurable benchmark regarding stochastic partial observability. 
We evaluate AERIAL in Dec-Tiger as well as in a variety of SMAC and MessySMAC maps,  and compare the results with state-based CTDE.  Furthermore,  we evaluate the robustness of AERIAL and state-based CTDE against various stochasticity configurations in MessySMAC.
\end{abstract}

\section{Introduction}\label{sec:introduction}

A wide range of real-world applications like fleet management,  industry 4.0, or communication networks can be formulated as \emph{decentralized partially observable Markov decision process (Dec-POMDP)} representing a cooperative \emph{multi-agent system (MAS)}, where multiple agents have to coordinate to achieve a common goal \cite{oliehoek2016concise}.  \emph{Stochastic partial observability} poses a major challenge for decentralized coordination in Dec-POMDPs due to noisy sensors and potentially high variance in initial states which are common in the real world \cite{kaelbling1998planning,oliehoek2016concise}.

\emph{Multi-agent reinforcement learning (MARL)} is a general approach to tackle Dec-POMDPs with remarkable progress in recent years \cite{wang2020qplex,wen2022multi}.  State-of-the-art MARL is based on \emph{centralized training for decentralized execution (CTDE)},  where training takes place in a laboratory or a simulator with access to global information \cite{lowe2017multi,foerster2017counterfactual}.  For example,  \emph{state-based CTDE} exploits true state information to learn a centralized value function in order to derive coordinated policies for decentralized decision making \cite{rashid2018qmix,yu2022the}.  Due to its effectiveness in the \emph{StarCraft Multi-Agent Challenge (SMAC)} as the current de facto standard for MARL evaluation,  state-based CTDE has become very popular and is widely considered an adequate approach to general Dec-POMDPs for more than half a decade,  leading to the development of many increasingly complex algorithms \cite{lyu2021contrasting,lyu2022deeper}.

However,  merely relying on state-based CTDE and SMAC in MARL research can be a pitfall in practice as stochastic partial observability is largely neglected -- despite being an important aspect in Dec-POMDPs \cite{lyu2022deeper}:

From an \emph{algorithm perspective},  purely state-based value functions are insufficient to evaluate and adapt multi-agent behavior,  since all agents make decisions on a completely different basis,  i.e.,  individual histories of noisy observations and actions.  True Dec-POMDP value functions consider more accurate closed-loop information about decentralized agent decisions though \cite{oliehoek2008optimal}.  Furthermore,  the optimal state-based value function represents an upper-bound of the true optimal Dec-POMDP value function thus state-based CTDE can result in overly optimistic behavior in general Dec-POMDPs \cite{lyu2022deeper}. 

From a \emph{benchmark perspective},  SMAC has very limited stochastic partial observability due to deterministic observations and low variance in initial states \cite{ellissmacv2}.  Therefore,  SMAC scenarios only represent simplified special cases rather than general Dec-POMDP challenges, being insufficient for assessing practicability of MARL.

In this paper,  we propose \emph{Attention-based Embeddings of Recurrence In multi-Agent Learning (AERIAL)} to approximate value functions under agent-wise stochastic partial observability.  AERIAL replaces the true state with a learned representation of multi-agent recurrence,  considering more accurate closed-loop information about decentralized agent decisions than state-based CTDE.  We then introduce \emph{MessySMAC},  a modified version of SMAC with stochastic observations and higher variance in initial states,  to provide a more general and configurable Dec-POMDP benchmark for more adequate evaluation.  Our contributions are as follows:
\begin{itemize}
\item We formulate and discuss the concepts of AERIAL w.r.t.  stochastic partial observability in Dec-POMDPs.
\item We introduce MessySMAC to enable systematic evaluation under various stochasticity configurations. 
\item We evaluate AERIAL in Dec-Tiger,  a small and traditional Dec-POMDP benchmark,  as well as in a variety of original SMAC and MessySMAC maps,  and compare the results with state-based CTDE.  Our results show that AERIAL achieves competitive performance in original SMAC,  and superior performance in Dec-Tiger and MessySMAC.  Furthermore,  we evaluate the robustness of AERIAL and state-based CTDE against various stochasticity configurations in MessySMAC.
\end{itemize}

\section{Background}\label{sec:background}

\subsection{Decentralized POMDPs}\label{subsec:dec_pomdps}

We formulate cooperative MAS problems as \emph{Dec-POMDP} $M = \langle \mathcal{D},\mathcal{S},\mathcal{A},\mathcal{T},\mathcal{R},\mathcal{Z},\Omega, b_{0} \rangle$, where $\mathcal{D} = \{1,...,N\}$ is a set of agents $i$, $\mathcal{S}$ is a set of (true) states $s_{t}$ at time step $t$, $\mathcal{A} = \langle \mathcal{A}_{i} \rangle_{i \in \mathcal{D}}$ is the set of joint actions $\mathbf{a_{t}} = \langle a_{t,1},...,a_{t,N} \rangle = \langle a_{t,i} \rangle_{i \in \mathcal{D}}$, $\mathcal{T}(s_{t+1}|s_{t}, \mathbf{a_{t}})$ is the state transition probability, $r_{t} = \mathcal{R}(s_{t},\mathbf{a_{t}}) \in \mathbb{R}$ is the shared reward,  $\mathcal{Z}$ is a set of local observations $z_{t,i}$ for each agent $i \in \mathcal{D}$,  $\Omega(\mathbf{z_{t+1}}|\mathbf{a_{t}},s_{t+1})$ is the probability of joint observation $\mathbf{z_{t+1}} = \langle z_{t+1,i} \rangle_{i \in \mathcal{D}} \in \mathcal{Z}^{N}$,  and $b_{0}$ is the probability distribution over initial states $s_{0}$ \cite{oliehoek2016concise}.  Each agent $i$ maintains a \emph{local history} $\tau_{t,i} \in (\mathcal{Z} \times \mathcal{A}_{i})^{t}$ and $\boldsymbol\tau_{\mathbf{t}} = \langle\tau_{t,i}\rangle_{i \in \mathcal{D}}$ is the \emph{joint history}. 
A \emph{belief state} $b(s_{t}|\boldsymbol\tau_{\mathbf{t}})$ is a sufficient statistic for joint history $\boldsymbol\tau_{\mathbf{t}}$ and defines a probability distribution over true states $s_{t}$,  updatable by Bayes' theorem \cite{kaelbling1998planning}.  Joint quantities are written in bold face.

Stochastic partial observability in $M$ is given by \emph{observation} and \emph{initialization stochasticity} w.r.t.  $\Omega$ and $b_{0}$ respectively.

A \emph{joint policy} $\boldsymbol\pi = \langle \pi_{i} \rangle_{i \in \mathcal{D}}$ with \emph{decentralized} or \emph{local policies} $\pi_i$ defines a deterministic mapping from joint histories to joint actions $\boldsymbol\pi(\boldsymbol\tau_{\mathbf{t}}) =  \langle \pi_{i}(\tau_{t,i}) \rangle_{i \in \mathcal{D}} \in \mathcal{A}$. 
The \emph{return} is defined by $G_{t} = \sum_{c=0}^{T-1} \gamma^{c} r_{t+c}$,  where $T$ is the \emph{horizon} and $\gamma \in [0,1]$ is the \emph{discount factor}.  $\boldsymbol\pi$ can be evaluated with a \emph{value function} $Q^{\boldsymbol\pi}(\boldsymbol\tau_{\mathbf{t}},\mathbf{a_{t}}) = \mathbb{E}_{b_{0},\mathcal{T},\Omega}[G_{t}|\boldsymbol\tau_{\mathbf{t}},\mathbf{a_{t}},\boldsymbol\pi]$. 
The goal is to find an \emph{optimal joint policy} $\boldsymbol\pi^{\mathbf{*}}$ with \emph{optimal value function} $Q^{\boldsymbol\pi^{\mathbf{*}}} = Q^{*}$ as defined in the next section.

\subsection{Optimal Value Functions and Policies}\label{subsec:optimal_value_functions}
\paragraph{\textbf{Fully Observable MAS}}
In MDP-like settings with a centralized controller,  the optimal value function $Q_{\textit{MDP}}^{*}$ is defined by \cite{watkins1992q,boutilier1996planning}:
\begin{equation}\label{eq:mdp_value_function}
Q_{\textit{MDP}}^{*}(s_{t},  \mathbf{a_{t}}) = r_{t} + \gamma \sum_{s_{t+1} \in \mathcal{S}} \mathcal{X}
\end{equation}

where $\mathcal{X} = \mathcal{T}(s_{t+1}|s_{t}, \mathbf{a_{t}}) \textit{max}_{\mathbf{a_{t+1}} \in \mathcal{A}} Q_{\textit{MDP}}^{*}(s_{t+1},  \mathbf{a_{t+1}})$.

Due to full observability,  $Q_{\textit{MDP}}^{*}$ does not depend on $\boldsymbol\tau_{\mathbf{t}}$ but on $s_{t}$.  Thus,  decentralized observations $z_{t,i}$ and probabilities according to $\Omega$ and $b_{0}$ are not considered at all.  An optimal (joint) policy $\boldsymbol\pi_{\textbf{MDP}}^{\mathbf{*}}$ of the centralized controller simply maximizes $Q_{\textit{MDP}}^{*}$ for all $s_{t}$ \cite{watkins1992q}:
\begin{equation}\label{eq:mdp_optimal_policy}
\boldsymbol\pi_{\textbf{MDP}}^{\mathbf{*}} = \textit{argmax}_{\boldsymbol\pi_{\textbf{MDP}}} \sum_{s_{t} \in \mathcal{S}} Q_{\textit{MDP}}^{*}(s_{t},  \boldsymbol\pi_{\textbf{MDP}}(s_{t}))
\end{equation}

\paragraph{\textbf{Partially Observable MAS}}
In general Dec-POMDPs,  where true states are not fully observable and only decentralized controllers or agents exist,  the optimal value function $Q^{*}$ is defined by \cite{oliehoek2008optimal}:
\begin{equation}\label{eq:dec_pomdp_value_function}
Q^{*}(\boldsymbol\tau_{\mathbf{t}},\mathbf{a_{t}}) = \sum_{s_{t} \in \mathcal{S}} b(s_{t}|\boldsymbol\tau_{\mathbf{t}}) \left(r_{t} + \\
\gamma \sum_{s_{t+1} \in \mathcal{S}}  \sum_{\mathbf{z_{t+1}} \in \mathcal{Z}^N} \mathcal{X} \right)
\end{equation}

where $\mathcal{X} = \mathcal{T}(s_{t+1}|s_{t}, \mathbf{a_{t}}) \Omega(\mathbf{z_{t+1}}|\mathbf{a_{t}},s_{t+1}) Q^{*}(\boldsymbol\tau_{\mathbf{t+1}}, $ $\boldsymbol\pi^{\mathbf{*}}(\boldsymbol\tau_{\mathbf{t+1}}))$ with $\boldsymbol\tau_{\mathbf{t+1}} = \langle \boldsymbol\tau_{\mathbf{t}},  \mathbf{a_{t}},  \mathbf{z_{t+1}} \rangle$.

An optimal joint policy $\boldsymbol\pi^{\mathbf{*}}$ for decentralized execution maximizes the expectation of $Q^{*}$ for all joint histories $\boldsymbol\tau_{\mathbf{t}}$ \cite{emery2004approximate,oliehoek2008optimal}:
\begin{equation}\label{eq:decision_rule}
\boldsymbol\pi^{\mathbf{*}} = \textit{argmax}_{\boldsymbol\pi} \sum_{t=0}^{T-1} \sum_{\boldsymbol\tau_{\mathbf{t}}  \in (\mathcal{Z}^N \times \mathcal{A})^{t}} \mathcal{C}^{\boldsymbol\pi}(\boldsymbol\tau_{\mathbf{t}})  \mathbf{P}^{\boldsymbol\pi}(\boldsymbol\tau_{\mathbf{t}}|b_{0})Q^{*}(\cdot)
\end{equation}
where $Q^{*}(\cdot) = Q^{*}(\boldsymbol\tau_{\mathbf{t}},  \boldsymbol\pi(\boldsymbol\tau_{\mathbf{t}}))$,  indicator $\mathcal{C}^{\boldsymbol\pi}(\boldsymbol\tau_{\mathbf{t}})$ filters out joint histories $\boldsymbol\tau_{\mathbf{t}}$ that are inconsistent with $\boldsymbol\pi$,  and probability $\mathbf{P}^{\boldsymbol\pi}(\boldsymbol\tau_{\mathbf{t}}|b_{0})$ represents the \emph{recurrence} of all agents considering agent-wise stochastic partial observability w.r.t.  decentralization of $\boldsymbol\pi$ and $\boldsymbol\tau_{\mathbf{t}}$ \cite{oliehoek2008optimal}:
\begin{align}\label{eq:history_probability}
\begin{split}
\mathbf{P}^{\boldsymbol\pi}(\boldsymbol\tau_{\mathbf{t}}|b_{0}) &= \mathbf{P}(\mathbf{z_{0}}|b_{0}) \prod\nolimits^{t}_{c=1} \mathbf{P}(\mathbf{z_{c}}|\boldsymbol\tau_{\mathbf{c-1}},\boldsymbol\pi)\\
																								&= \mathbf{P}(\mathbf{z_{0}}|b_{0}) \prod\nolimits^{t}_{c=1} \sum_{s_{c} \in \mathcal{S}} \sum_{s_{c-1} \in \mathcal{S}} \mathcal{T}(\cdot) \Omega(\cdot)
\end{split}
\end{align}
where $\mathcal{T}(\cdot) = \mathcal{T}(s_{c}|s_{c-1}, \boldsymbol\pi(\boldsymbol\tau_{\mathbf{c-1}}))$ and $\Omega(\cdot) = \Omega(\mathbf{z_{c}}|\boldsymbol\pi(\boldsymbol\tau_{\mathbf{c-1}}),s_{c})$.

Since all agents act according to their local history $\tau_{t,i}$ without access to the complete joint history $\boldsymbol\tau_{\mathbf{t}}$,  recurrence $\mathbf{P}^{\boldsymbol\pi}(\boldsymbol\tau_{\mathbf{t}}|b_{0})$ depends on more accurate \emph{closed-loop information} than just true states $s_t$,  i.e.,  all previous observations,  actions,  and probabilities according to $b_{0}$,  $\mathcal{T}$,  and $\Omega$.

$Q_{\textit{MDP}}^{*}$ is proven to represent an upper bound of $Q^{*}$ \cite{oliehoek2008optimal}.  Thus,  naively deriving local policies $\pi_{i}$ from $Q_{\textit{MDP}}^{*}$ instead of $Q^{*}$ can result in overly optimistic behavior as we will show in Section \ref{subsec:state_vs_history} and \ref{sec:experiments}.

\subsection{Multi-Agent Reinforcement Learning}\label{subsec:marl}

Finding an optimal joint policy $\boldsymbol\pi^{\mathbf{*}}$ via exhaustive computation of $Q^{*}$ according to Eq. \ref{eq:dec_pomdp_value_function}-\ref{eq:history_probability} is intractable in practice \cite{nair2003taming,szer2005maa}.  MARL offers a scalable way to learn $Q^{*}$ and $\boldsymbol\pi^{\mathbf{*}}$ via function approximation,  e.g.,  using CTDE,  where training takes place in a laboratory or a simulator with access to global information \cite{lowe2017multi,foerster2017counterfactual}.  We focus on value-based MARL to learn a centralized value function $Q_{\textit{tot}} \approx Q^{*}$,  which can be factorized into \emph{local utility functions} $\langle Q_{i} \rangle_{i \in \mathcal{D}}$ for decentralized decision making via $\pi_{i}(\tau_{t,i}) = \textit{argmax}_{a_{t,i}} Q_{i}(\tau_{t,i}, a_{t,i})$. 
For that,  a \emph{factorization operator} $\Psi$ is used \cite{phan2021vast}:
\begin{equation}\label{eq:vff}
Q_{\textit{tot}}(\boldsymbol\tau_{\mathbf{t}}, \mathbf{a_{t}}) = \Psi(Q_{1}(\tau_{t,1},a_{t,1}),...,Q_{N}(\tau_{t,N},a_{t,N}))
\end{equation}
In practice, $\Psi$ is realized with deep neural networks, such that $\langle Q_{i} \rangle_{i \in \mathcal{D}}$ can be learned end-to-end via backpropagation by minimizing the mean squared \emph{temporal difference (TD)} error \cite{rashid2018qmix,sunehag2017value}.  A factorization operator $\Psi$ is \emph{decentralizable} when satisfying the \emph{IGM (Individual-Global-Max)} such that \cite{son2019qtran}:

\begin{equation}\label{eq:igm}
\textit{argmax}_{\mathbf{a_{t}}}Q_{\textit{tot}}(\boldsymbol\tau_{\mathbf{t}}, \mathbf{a_{t}}) =
\begin{pmatrix} 
\textit{argmax}_{a_{t,1}}Q_{1}(\tau_{t,1},a_{t,1})\\
\vdots \\
\textit{argmax}_{a_{t,N}}Q_{N}(\tau_{t,N},a_{t,N})
\end{pmatrix}
\end{equation}

There exists a variety of factorization operators $\Psi$ which satisfy Eq.  \ref{eq:igm} using monotonicity like QMIX \cite{rashid2018qmix},  nonlinear transformation like QPLEX \cite{wang2020qplex},  or loss weighting like CW- and OW-QMIX \cite{rashid2020weighted}.  Most approaches use state-based CTDE to learn $Q_{\textit{MDP}}^{*}$ according to Eq.  \ref{eq:mdp_value_function} instead of $Q^{*}$ (Eq.  \ref{eq:dec_pomdp_value_function}-\ref{eq:history_probability}).

\subsection{Recurrent Reinforcement Learning}\label{subsec:recurrent_rl}

In partially observable settings,  the policy $\pi_{i}$ of agent $i$ conditions on the history $\tau_{t,i}$ of past observations and actions \cite{kaelbling1998planning,oliehoek2016concise}.  In practice,  \emph{recurrent neural networks (RNNs)} like LSTMs or GRUs are used to learn a compact representation $h_{t,i}$ of $\tau_{t,i}$ and $\pi_{i}$ known as \emph{hidden state} or \emph{memory representation}\footnote{In this paper,  we use the term \emph{memory representation} to avoid confusion with the state terminology of the (Dec-)POMDP literature \cite{kaelbling1998planning,oliehoek2016concise}.},  which implicitly encodes the \emph{individual recurrence} of agent $i$,  i.e.,  the distribution $P_{i}^{\pi_{i}}$ over $\tau_{t,i}$ \cite{hochreiter1997long,cho2014properties,hu2019simplified}:
\begin{align}\label{eq:recurrent_history_probability}
P_{i}^{\pi_{i}}(\tau_{t,i}|b_{0}) = P_{i}(z_{0,i}|b_{0}) \prod\nolimits^{t}_{c=1} P_{i}(z_{c,i}|\tau_{c-1,i}, \pi_{i})
\end{align}
RNNs are commonly used for partially observable problems and have been empirically shown to be more effective than using raw observations $z_{t,i}$ or histories $\tau_{t,i}$ \cite{hausknecht2015deep,samvelyan2019starcraft,vinyals2019grandmaster}.

\section{Related Work}\label{sec:related_work}

\paragraph{\textbf{Multi-Agent Reinforcement Learning}}

In recent years,  MARL has achieved remarkable progress in challenging domains \cite{gupta2017cooperative,vinyals2019grandmaster}.  State-of-the-art MARL is based on CTDE to learn a centralized value function $Q_{\textit{tot}}$ for actor-critic learning \cite{lowe2017multi,foerster2017counterfactual,yu2022the} or factorization \cite{rashid2018qmix,rashid2020weighted,wang2020qplex}.  However,  the majority of works assumes a simplified Dec-POMDP setting,  where $\Omega$ is deterministic,  and uses true states to approximate $Q_{\textit{MDP}}^{*}$ according to Eq.  \ref{eq:mdp_value_function} instead of $Q^{*}$ (Eq.  \ref{eq:dec_pomdp_value_function}-\ref{eq:history_probability}).  Thus, state-based CTDE is possibly less effective in more general Dec-POMDP settings. 
Our approach addresses stochastic partial observability with a \emph{learned representation} of multi-agent recurrence $\mathbf{P}^{\boldsymbol\pi}(\boldsymbol\tau_{\mathbf{t}}|b_{0})$ according to Eq.  \ref{eq:history_probability} instead of $s_t$. 

\paragraph{\textbf{Weaknesses of State-Based CTDE}}

Recent works investigated potential weaknesses of state-based CTDE for multi-agent actor-critic methods regarding bias and variance \cite{lyu2021contrasting,lyu2022deeper}.  The experimental results show that state-based CTDE can surprisingly fail in very simple Dec-POMDP benchmarks that exhibit more stochasticity than SMAC.  While these studies can be considered an important step towards general Dec-POMDPs,  there is neither an approach which adequately addresses stochastic partial observability nor a benchmark to systematically evaluate such an approach yet.
In this work,  we focus on \emph{value-based MARL},  where learning an accurate value function is important for meaningful factorization,  and propose an \emph{attention-based recurrence approach} to approximate value functions under stochastic partial observability.  We also introduce a \emph{modified} SMAC benchmark,  which enables systematic evaluation under various stochasticity configurations. 

\paragraph{\textbf{Attention-Based CTDE}}

Attention has been used in CTDE to process information of potentially variable length $N$,  where joint observations $\mathbf{z_{t}}$,  joint actions $\mathbf{a_{t}}$,  or local utilities $\langle Q_{i} \rangle_{i \in \mathcal{D}}$ are weighted and aggregated to provide a meaningful representation for value function approximation \cite{iqbal19a,wang2020qplex,iqbal2021randomized,wen2022multi,khan2022transformer}. Most works focus on Markov games without observation stochasticity, which are special cases of the Dec-POMDP setting. 
In this work,  we focus on \emph{stochastic partial observability} and apply \emph{self-attention} to the \emph{memory representations} $h_{t,i}$ of all agents' RNNs instead of the raw observations $z_{t,i}$ to approximate $Q^{*}$ for \emph{general Dec-POMDPs} according to Eq. \ref{eq:dec_pomdp_value_function}-\ref{eq:history_probability}.

\begin{figure*}
  \centering
    \includegraphics[width=0.9\textwidth]{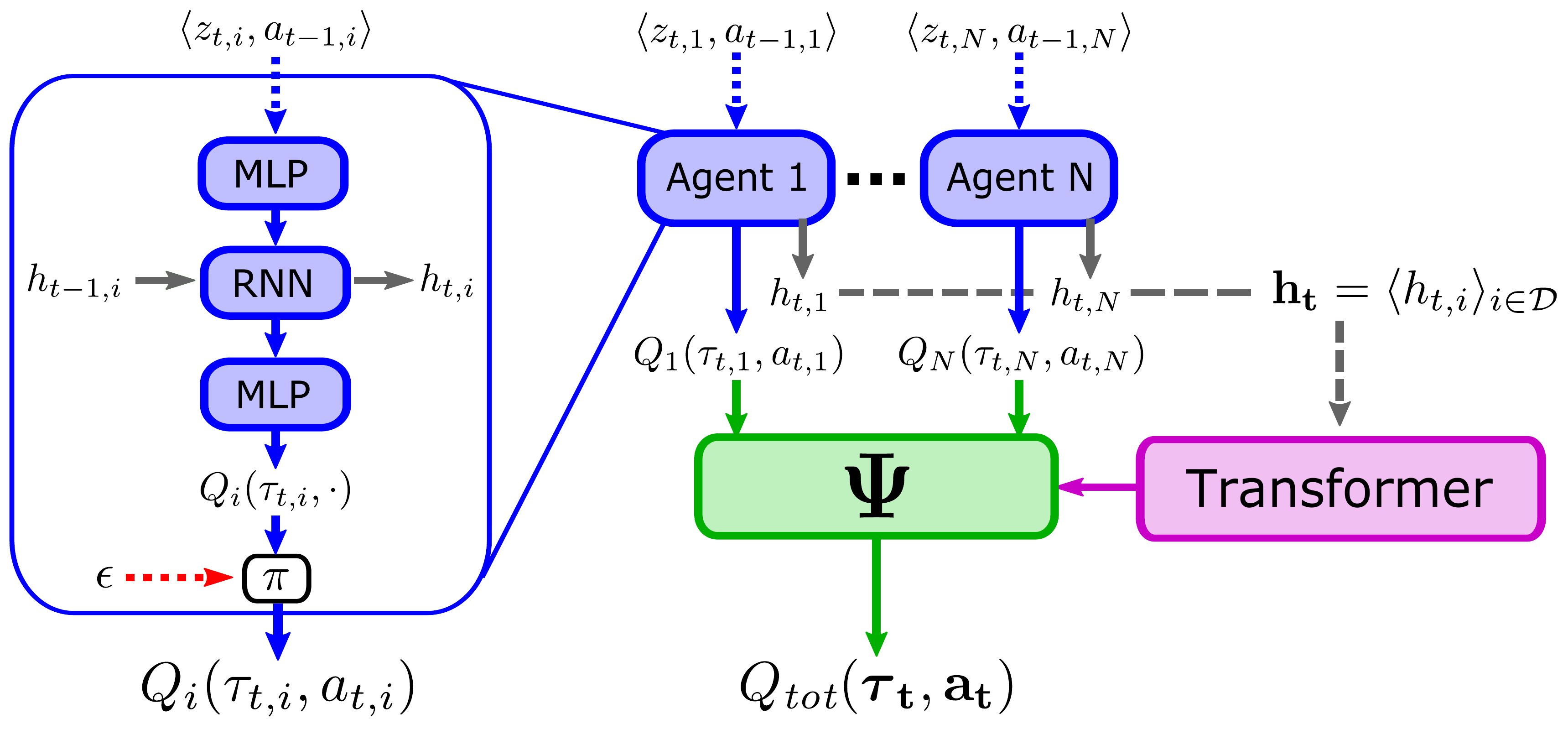}
    \caption{Illustration of the AERIAL setup.  \textit{Left:} Recurrent agent network structure with memory representations $h_{t-1,i}$ and $h_{t,i}$.  \textit{Right:} Value function factorization via factorization operator $\Psi$ using the joint memory representation $\mathbf{h_{t}} =  \langle h_{t,i} \rangle_{i \in \mathcal{D}}$ of all agents' RNNs instead of true states $s_{t}$.  All memory representations $h_{t,i}$ are detached from the computation graph to avoid additional differentiation (indicated by the dashed gray arrows) and passed through a simplified transformer before being used by $\Psi$ for value function factorization. }
    \label{fig:aerial_architecture}
\end{figure*}

\section{AERIAL}

\subsection{Limitation of State-Based CTDE}\label{subsec:state_vs_history}

Most state-of-the-art works assume a simplified Dec-POMDP setting,  where $\Omega$ is deterministic,  and approximate $Q_{\textit{MDP}}^{*}$ according to Eq.  \ref{eq:mdp_value_function} instead of $Q^{*}$ (Eq.  \ref{eq:dec_pomdp_value_function}-\ref{eq:history_probability}).

If there are only deterministic observations and initial states $s_{\textit{0}}$ such that  $b_{0}(s_{\textit{0}}) = 1$ and $b_{0}(s') = 0$ if $s' \neq s_{\textit{0}}$, then multi-agent recurrence $\mathbf{P}^{\boldsymbol\pi}(\boldsymbol\tau_{\mathbf{t}}|b_{0})$ as defined in Eq.  \ref{eq:history_probability} would only depend on state transition probabilities $\mathcal{T}(s_{t+1}|s_{t}, \mathbf{a_{t}})$ which are purely state-based,  ignoring decentralization of agents and observations \cite{oliehoek2008optimal}. In such scenarios, stochastic partial observability is very limited,  especially if all $\pi_{i}$ are deterministic. 
We hypothesize that this is one reason for the empirical success of state-based CTDE in original SMAC,  whose scenarios seemingly have these simplifying properties \cite{ellissmacv2}.

In the following, we regard a small example, where state-based CTDE can fail at finding an optimal joint policy $\boldsymbol\pi^{\mathbf{*}}$.

\paragraph{\textbf{Example}} \emph{Dec-Tiger} is a traditional and simple Dec-POMDP benchmark with $N = 2$ agents facing two doors \cite{nair2003taming}. A tiger is randomly placed behind the left ($s_{L}$) or right door ($s_{R}$) representing the true state.  Both agents are able to listen ($\textit{li}$) and open the left ($o_{L}$) or right door ($o_{R}$).  The listening action $\textit{li}$ produces a noisy observation of either hearing the tiger to be left ($z_{L}$) or right ($z_{R}$), which correctly indicates the tiger's position with $85\%$ chance and a cost of $-1$ per listening agent.  If both agents open the same door,  the episode terminates with a reward of -50 if opening the tiger door and +20 otherwise.  If both agents open different doors,  the episode ends with -100 reward and,  if only one agent opens a door while the other agent is listening,  the episode terminates with -101 if opening the tiger door and +9 otherwise.

Given a horizon of $T=2$,  the tiger being behind the right door ($s_{R}$),  and both agents having listened in the first step,  where agent $1$ heard $z_{L}$ and agent $2$ heard $z_{R}$: Assuming that both agents learned to perform the same actions,  e.g.,  due to CTDE and parameter sharing \cite{tan1993multi,gupta2017cooperative},  $Q_{\textit{MDP}}^{*}$ and $Q^{*}$ would estimate the following values\footnote{The exact calculation is provided in the Appendix \ref{subsec:appendix_dec_tiger}.}:
\begin{align*}
&Q_{\textit{MDP}}^{*}(s_{R},  \langle \textit{li},  \textit{li} \rangle) = -2 & &\color{blue}{Q^{*}(\boldsymbol\tau_{\mathbf{t}},  \langle \textit{li},  \textit{li} \rangle) = -2}\\
&\color{blue}{Q_{\textit{MDP}}^{*}(s_{R},  \langle o_{L},  o_{L} \rangle) = 20} & &Q^{*}(\boldsymbol\tau_{\mathbf{t}},  \langle o_{L},  o_{L} \rangle) = -15\\
&Q_{\textit{MDP}}^{*}(s_{R},   \langle o_{R},  o_{R} \rangle) = -50 & &Q^{*}(\boldsymbol\tau_{\mathbf{t}},  \langle o_{R},  o_{R} \rangle) = -15
\end{align*}

Any policy $\boldsymbol\pi_{\textbf{MDP}}^{\mathbf{*}}$ or decentralizable joint policy $\boldsymbol\pi$ w.r.t.  IGM (Eq.  \ref{eq:igm}) that maximizes $Q_{\textit{MDP}}^{*}$ according to Eq.  \ref{eq:mdp_optimal_policy} would optimistically recommend $\langle o_{L},  o_{L} \rangle$ based on the true state $s_{R}$,  regardless of what the agents observed.  However,  any joint policy $\boldsymbol\pi^{\mathbf{*}}$ that maximizes the expectation of $Q^{*}$ according to Eq.  \ref{eq:decision_rule} would consider agent-wise stochastic partial observability and recommend $\langle \textit{li},  \textit{li} \rangle$,  which corresponds to the true optimal decision for $T = 2$ \cite{szer2005maa}. 

\subsection{Attention-Based Embeddings of Recurrence}\label{subsec:aerial_concept}
\paragraph{\textbf{Preliminaries}}
We now introduce \emph{Attention-based Embeddings of Recurrence In multi-Agent Learning (AERIAL)} to approximate optimal Dec-POMDP value functions $Q^{*}$ according to Eq. \ref{eq:dec_pomdp_value_function}-\ref{eq:history_probability}.  Our setup uses a factorization operator $\Psi$ like QMIX or QPLEX according to Eq. \ref{eq:vff}-\ref{eq:igm}.  All agents process their local histories $\tau_{t,i}$ via RNNs as motivated in Section \ref{subsec:recurrent_rl} and schematically shown in Fig.  \ref{fig:aerial_architecture} (left). 

Unlike $Q_{\textit{MDP}}^{*}$,  the true optimal Dec-POMDP value function $Q^{*}$ considers more accurate closed-loop information about decentralized agent decisions through multi-agent recurrence $\mathbf{P}^{\boldsymbol\pi}(\boldsymbol\tau_{\mathbf{t}}|b_{0})$ according to Eq.  \ref{eq:history_probability}.
Simply replacing $s_{t}$ with $\boldsymbol\tau_{\mathbf{t}}$ as suggested in \cite{lyu2022deeper} is not sufficient because the resulting value function would assume a centralized controller with access to the complete joint history $\boldsymbol\tau_{\mathbf{t}}$,  in contrast to decentralized agents $i$ which can only access their respective local history $\tau_{t,i}$ \cite{oliehoek2008optimal}. 

\paragraph{\textbf{Exploiting Multi-Agent Recurrence}}
At first we propose to naively exploit all individual recurrences by simply replacing the true state $s_{t}$ in CTDE with the \emph{joint memory representation} $\mathbf{h_{t}} =  \langle h_{t,i} \rangle_{i \in \mathcal{D}}$ of all agents' RNNs.  Each memory representation $h_{t,i}$ implicitly encodes the individual recurrence $P_{i}^{\pi_{i}}(\tau_{t,i}|b_{0})$ of agent $i$ according to Eq. \ref{eq:recurrent_history_probability}.  Therefore,  $\mathbf{h_{t}}$ provides more accurate closed-loop information about decentralized agent decisions than $s_t$.

This approach,  called \texttt{AERIAL (no attention)}, can already be considered a sufficient solution if all individual recurrences $P_{i}^{\pi_{i}}(\tau_{t,i}|b_{0})$ are statistically independent such that $\mathbf{P}^{\boldsymbol\pi}(\boldsymbol\tau_{\mathbf{t}}|b_{0}) = \prod^{N}_{i = 1} P_{i}^{\pi_{i}}(\tau_{t,i}|b_{0})$.

\paragraph{\textbf{Attention-Based Recurrence}}
While \texttt{AERIAL (no attention)} offers a simple way to address agent-wise stochastic partial observability,  the independence assumption of all individual recurrences $P_{i}^{\pi_{i}}(\tau_{t,i}|b_{0})$ does not hold in practice due to correlations in observations and actions \cite{bernstein2005bounded,amato2007optimizing}. 

Given the Dec-Tiger example above, the individual recurrences according to Eq.  \ref{eq:recurrent_history_probability} are $P_{1}^{\pi_{1}}(\tau_{t,1}|b_{0}) = P_{2}^{\pi_{2}}(\tau_{t,2}|b_{0}) = 0.5$ \cite{kaelbling1998planning}.  However,  the actual multi-agent recurrence according to Eq.  \ref{eq:history_probability} is $\mathbf{P}^{\boldsymbol\pi}(\boldsymbol\tau_{\mathbf{t}}|b_{0}) = 0.15\cdot0.85 \neq P_{1}^{\pi_{1}}(\tau_{t,1}|b_{0}) \cdot P_{2}^{\pi_{2}}(\tau_{t,2}|b_{0})$, indicating that individual recurrences are not statistically independent in general \cite{oliehoek2016concise}.

Therefore,  we process $\mathbf{h_{t}}$ by a simplified \emph{transformer} along the agent axis to automatically consider the latent dependencies of all memory representations $h_{t,i} \in \mathbf{h_{t}}$ through self-attention. The resulting approach, called \texttt{AERIAL}, is depicted in Fig. \ref{fig:aerial_architecture} and Algorithm \ref{algorithm:AERIAL} in Appendix \ref{sec:appendix_aerial_full_algorithm}.

Our transformer does not use positional encoding or masking, since we assume no particular ordering among agents. The joint memory representation $\mathbf{h_{t}}$ is passed through a single \emph{multi-head attention} layer with the output of each attention head $c$ being defined by \cite{vaswani2017attention}:
\begin{equation}\label{eq:self_attention}
\textit{att}_{c}(\mathbf{h_{t}}) = \textit{softmax}\left(\frac{W_{q}^{c}(\mathbf{h_{t}})W_{k}^{c}(\mathbf{h_{t}})^{\top}}{\sqrt{d_{\textit{att}}}}\right)W_{v}^{c}(\mathbf{h_{t}})
\end{equation}
where $W_{q}^{c}$,  $W_{k}^{c}$,  and $W_{v}^{c}$ are \emph{multi-layer perceptrons (MLP)} with an output dimensionality of $d_{\textit{att}}$.  All outputs $\textit{att}_{c}(\mathbf{h_{t}})$ are summed and passed through a series of MLP layers before being fed into the factorization operator $\Psi$,  effectively replacing the true state $s_{t}$ by a learned representation of multi-agent recurrence $\mathbf{P}^{\boldsymbol\pi}(\boldsymbol\tau_{\mathbf{t}}|b_{0})$ according to Eq.  \ref{eq:history_probability}.

To avoid additional differentation of $\mathbf{h_{t}}$ through $\Psi$ or Eq.  \ref{eq:self_attention},  we detach $\mathbf{h_{t}}$ from the computation graph. Thus,  we make sure that $\mathbf{h_{t}}$ is only learned through agent RNNs.

\subsection{Discussion of AERIAL}

The strong focus on state-based CTDE in the last few years has led to the development of increasingly complex algorithms that largely neglect stochastic partial observability in general Dec-POMDPs \cite{lyu2021contrasting,lyu2022deeper}.  In contrast,  AERIAL offers a simple way to adjust factorization approaches by replacing the true state $s_{t}$ with a learned representation of multi-agent recurrence $\mathbf{P}^{\boldsymbol\pi}(\boldsymbol\tau_{\mathbf{t}}|b_{0})$ to consider more accurate closed-loop information about decentralized agent decisions.  The rest of the training scheme remains unchanged,  which eases adjustment of existing approaches. 

Since the naive independence assumption of individual memory representations $h_{t,i}$ does not hold in practice -- despite decentralization -- we use a simplified transformer to consider the latent dependencies of all $h_{t,i} \in \mathbf{h_{t}}$ along the agent axis to learn an adequate representation of multi-agent recurrence $\mathbf{P}^{\boldsymbol\pi}(\boldsymbol\tau_{\mathbf{t}}|b_{0})$ according to Eq.  \ref{eq:history_probability}.

AERIAL does not depend on true states therefore requiring less overall information than state-based CTDE,  since we assume $\mathbf{h_{t}}$ to be available in all CTDE setups anyway \cite{foerster2017counterfactual,rashid2020weighted}.  Note that AERIAL does not necessarily require RNNs to obtain $\mathbf{h_{t}}$ as hidden layers of MLPs or decision transformers can be used to approximate $\mathbf{h_{t}}$ as well \cite{son2019qtran,chen2021decision}.

\begin{figure*}
\centering
     \subfigure{%
         \includegraphics[width=0.75\textwidth]{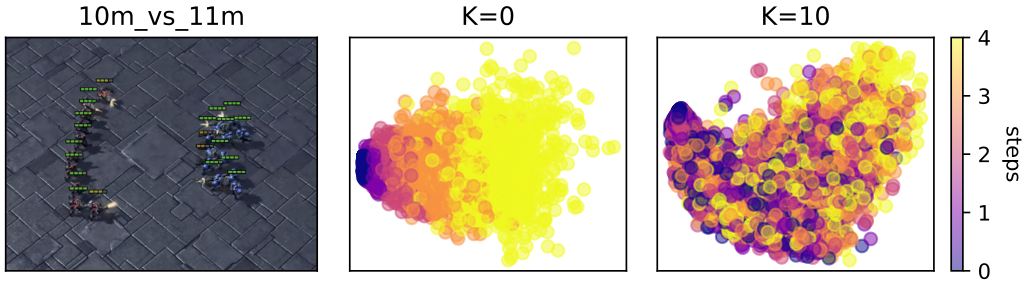}
     }\\
     \subfigure{%
         \includegraphics[width=0.75\textwidth]{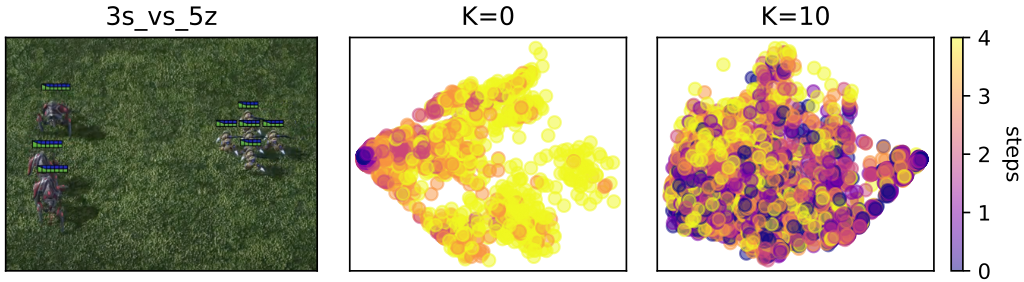}
     }
     \caption{\textit{Left:} Screenshot of two SMAC maps.   \textit{Middle:} PCA visualization of the joint observations in original SMAC within the first 5 steps of 1,000 episodes using a random policy with $K=0$ initial random steps.  \textit{Right:} Analogous PCA visualization for MessySMAC with $K=10$ initial random steps.  For visual comparability,  the observations are deterministic here.}
     \label{fig:initial_state_visualization}
\end{figure*}
\newpage
\section{MessySMAC}\label{sec:messy_smac_benchmark}

\subsection{Limitation of SMAC as a Benchmark}

\emph{StarCraft Multi-Agent Challenge (SMAC)} provides a rich set of micromanagement tasks,  where a team of learning agents has to fight against an enemy team,  which acts according to handcrafted heuristics of the built-in StarCraft AI \cite{samvelyan2019starcraft}.  SMAC currently represents the de facto standard for MARL evaluation \cite{rashid2018qmix,rashid2020weighted,wang2020qplex}.  However,  SMAC scenarios exhibit very limited stochastic partial observability due to deterministic observations and low variance in initial states therefore only representing simplified special cases rather than general Dec-POMDP challenges \cite{lyu2022deeper,ellissmacv2}.  To assess practicability of MARL,  we need benchmarks with sufficient stochasticity as the real-world is generally messy and only observable through noisy sensors.

\subsection{SMAC with Stochastic Partial Observability}\label{subsec:messy_smac}

\emph{MessySMAC} is a modified version of SMAC with \emph{observation stochasticity} w.r.t.  $\Omega$,  where all measured values of observation $z_{t,i}$ are negated with a probability of $\phi \in [0,1)$,  and \emph{initialization stochasticity} w.r.t.  $b_{0}$,  where $K$ random steps are initially performed before officially starting an episode.  During the initial phase,  the agents can already be ambushed by the built-in AI,  which further increases difficulty compared to the original SMAC maps if $K > 0$.  MessySMAC represents a more general Dec-POMDP challenge which enables systematic evaluation under various stochasticity configurations according to $\phi$ and $K$. 

Fig.  \ref{fig:initial_state_visualization} shows the PCA visualization of joint observations in two maps of original SMAC ($K = 0$) and MessySMAC ($K = 10$) within the first 5 steps of 1,000 episodes using a random policy.  In original SMAC,  the initial observations of $s_{0}$ (dark purple) are very similar and can be easily distinguished from subsequent observations by merely regarding time steps.  Therefore,  open-loop control might already be sufficient to solve these scenarios satisfactorily as hypothesized in \cite{ellissmacv2}.  However,  the distinction of observations by time steps is more tricky in MessySMAC due to significantly higher entropy in $b_{0}$,  indicating higher initialization stochasticity and a stronger requirement for closed-loop control,  where agents need to explicitly consider their actual observations to make proper decisions. 

\subsection{Comparison with SMACv2}\label{subsec:comparison_smacv2}
\emph{SMACv2} is an update to the original SMAC benchmark featuring initialization stochasticity w.r.t.  position and unit types,  as well as observation restrictions \cite{ellissmacv2}. 
SMACv2 addresses similar issues as MessySMAC but MessySMAC additionally features \emph{observation stochasticity} w.r.t.  $\Omega$ according to the general Dec-POMDP formulation in Section \ref{subsec:dec_pomdps}.  Unlike MessySMAC,  SMACv2 does not support the \emph{original SMAC maps} thus not enabling direct comparability w.r.t.  stochasticity configurations. 

Therefore,  SMACv2 can be viewed as entirely new StarCraft II benchmark,  while MessySMAC represents a \emph{SMAC extension},  enabling systematic evaluation under various stochasticity configurations for the original SMAC maps.

\section{Experiments}\label{sec:experiments}

We use the state-based CTDE implementations of \texttt{QPLEX},  \texttt{CW-QMIX},  \texttt{OW-QMIX},  and \texttt{QMIX} from \cite{rashid2020weighted} as state-of-the-art baselines with their default hyperparameters.  We also integrate \texttt{MAPPO} from \cite{yu2022the}.  For all experiments,  we report the average performance and the 95\% confidence interval over at least 20 runs.

\texttt{AERIAL} is implemented\footnote{Code is available at {\color{blue}\url{https://github.com/thomyphan/messy_smac}}.  Further details are in Appendix \ref{subsec:appendix_technical}.} using \texttt{QMIX} as factorization operator $\Psi$ according to Fig.  \ref{fig:aerial_architecture}.  We also experimented with \texttt{QPLEX} as alternative with no significant difference in performance.  Thus,  we stick with \texttt{QMIX} for efficiency due to fewer trainable parameters.  The transformer of \texttt{AERIAL} has 4 heads with $W_{q}^{c}$,  $W_{k}^{c}$,  and $W_{v}^{c}$ each having one hidden layer of $d_{\textit{att}} = 64$ units with ReLU activation.  The subsequent MLP layers have 64 units with ReLU activation.

For ablation study,  we implement \texttt{AERIAL (no attention)},  which trains $\Psi$ directly on $\mathbf{h_{t}}$ without self-attention as described in Section \ref{subsec:aerial_concept},  and \texttt{AERIAL (raw history)},  which trains $\Psi$ on the raw joint history $\boldsymbol\tau_{\mathbf{t}}$ concatenated with the true state $s_{t}$ as originally proposed for actor-critic methods \cite{lyu2022deeper}.

\subsection{Dec-Tiger}\label{subsec:dec_tiger_results}

\paragraph{\textbf{Setting}}
We use the Dec-Tiger problem described in Section \ref{subsec:state_vs_history} and \cite{nair2003taming} as simple proof-of-concept domain with $T = 4$ and $\gamma = 1$.  We also provide the optimal value of 4.8 computed with MAA* \cite{szer2005maa}. 

\begin{figure}
  \centering
    \includegraphics[width=0.47\textwidth]{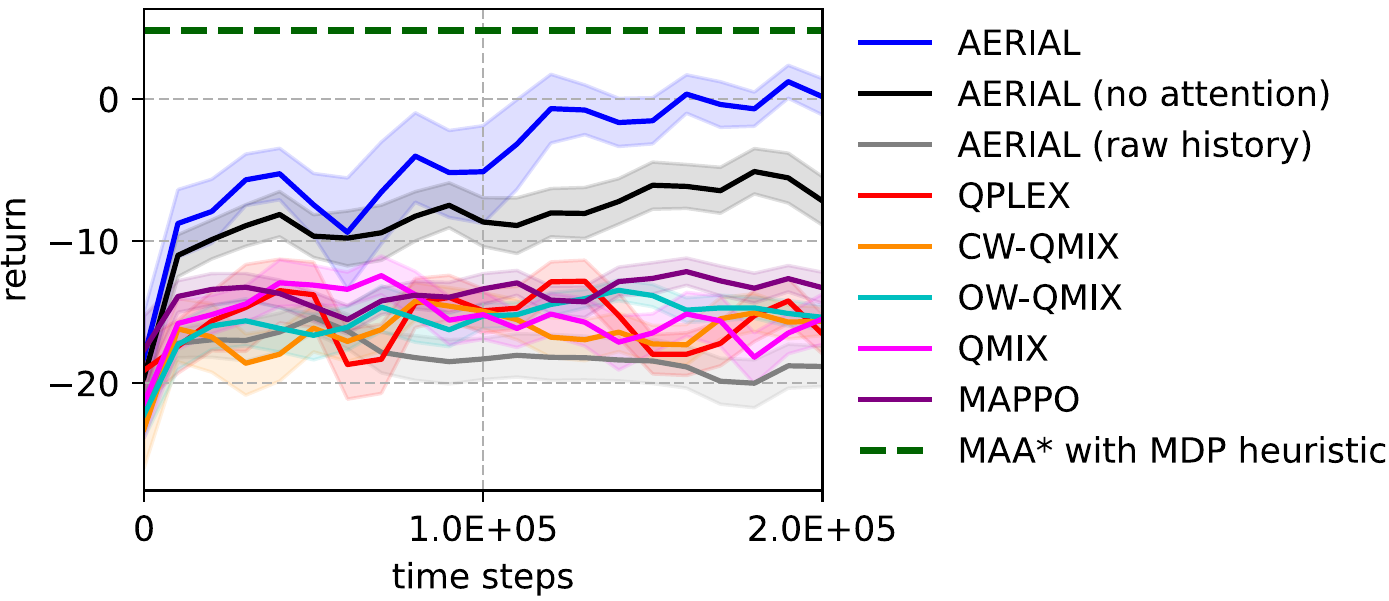}
    \caption{Average learning progress w.r.t.  the return of \texttt{AERIAL} variants and state-of-the-art baselines in Dec-Tiger over 50 runs.  Shaded areas show the 95\% confidence interval.}
    \label{fig:dec_tiger_results}
\end{figure}

\paragraph{\textbf{Results}}
The results are shown in Fig.  \ref{fig:dec_tiger_results}.  \texttt{AERIAL} comes closest to the optimum,  achieving an average return of about zero.  \texttt{AERIAL (no attention)} performs second best with an average return of about -8,  while all other approaches achieve an average return of about -15. 

\paragraph{\textbf{Discussion}}
The results confirm the example from Section \ref{subsec:state_vs_history} and the findings of \cite{oliehoek2008optimal,lyu2022deeper}.  All state-based CTDE approaches and \texttt{AERIAL (raw history)} converge to a one-step policy,  where both agents optimistically open the same door regardless of any agent observation.  \texttt{AERIAL (no attention)} converges to a local optimum most of the time,  where both agents only listen for all $T = 4$ time steps.  \texttt{AERIAL} performs best due to considering the latent dependencies of all memory representations $h_{t,i} \in \mathbf{h_{t}}$ via self-attention to learn an adequate representation of multi-agent recurrence $\mathbf{P}^{\boldsymbol\pi}(\boldsymbol\tau_{\mathbf{t}}|b_{0})$ according to Eq.  \ref{eq:history_probability}.

\subsection{Original SMAC }\label{subsec:smac_results}

\paragraph{\textbf{Setting}}
We evaluate \texttt{AERIAL} in original SMAC using the maps \texttt{3s5z} and \texttt{10m\_vs\_11m},  which are classified as \emph{easy},  as well as the \emph{hard} maps \texttt{2c\_vs\_64zg},  \texttt{3s\_vs\_5z},  and \texttt{5m\_vs\_6m},  and the \emph{super hard} map \texttt{3s5z\_vs\_3s6z} \cite{samvelyan2019starcraft}. 

\paragraph{\textbf{Results}}
The final average test win rates after 2 million steps of training are shown in Table \ref{tab:smac_results}.  \texttt{AERIAL} is competitive to \texttt{QPLEX} and \texttt{QMIX} in the easy maps,  while performing best in \texttt{3s\_vs\_5z} and \texttt{5m\_vs\_6m}.  \texttt{MAPPO} performs best in \texttt{2c\_vs\_64zg} and \texttt{3s5z\_vs\_3s6z} with \texttt{AERIAL} being second best in the super hard map \texttt{3s5z\_vs\_3s6z}.

\begin{table*}[!ht]
\centering\small
\caption{Average win rate of \texttt{AERIAL} and state-of-the-art baselines after 2 million time steps of training across 400 final test episodes for the original SMAC maps with the 95\% confidence interval.  The best results per map are highlighted in boldface and blue.}
\small
\begin{tabular}{|l||l|l|l|l|l|l|l|l|} \hline
 & \texttt{AERIAL} & \texttt{QPLEX} & \texttt{CW-QMIX} & \texttt{OW-QMIX} &  \texttt{QMIX} & \texttt{MAPPO}\\ \hline
\texttt{3s5z} & $\color{blue}\mathbf{0.95 \pm 0.01}$ & $0.94 \pm 0.01$ & $0.87 \pm 0.02$ & $0.91 \pm 0.02$  & $\color{blue}\mathbf{0.95 \pm 0.01}$ & $68.7 \pm 0.94$ \\
\texttt{10m\_vs\_11m} & $\color{blue}\mathbf{0.97 \pm 0.01}$ & $0.90 \pm 0.02$ & $0.91 \pm 0.02$ & $0.96 \pm 0.01$ & $0.90 \pm 0.02$ & $77.3 \pm 0.66$ \\
\texttt{2c\_vs\_64zg} & $0.52 \pm 0.11$ & $0.29 \pm 0.1$ & $0.38 \pm 0.12$ & $0.55 \pm 0.13$ & $0.59 \pm 0.11$ & $\color{blue}\mathbf{90.2 \pm 0.24}$\\
\texttt{3s\_vs\_5z} & $\color{blue}\mathbf{0.96 \pm 0.02}$ & $0.74 \pm 0.11$ & $0.18 \pm 0.06$ & $0.08 \pm 0.04$  & $0.81 \pm 0.05$ & $73.8 \pm 0.44$\\
\texttt{5m\_vs\_6m} & $\color{blue}\mathbf{0.77 \pm 0.03}$ & $0.66 \pm 0.04$ & $0.41 \pm 0.04$ & $0.55 \pm 0.06$  & $0.67 \pm 0.05$ & $60.6 \pm 1.13$\\
\texttt{3s5z\_vs\_3s6z} & $0.18 \pm 0.09$ & $0.1 \pm 0.03$ & $0.0 \pm 0.0$ & $0.02 \pm 0.01$ & $0.02 \pm 0.02$ & $\color{blue}\mathbf{20.5 \pm 2.91}$\\\hline
\end{tabular}\label{tab:smac_results}\normalsize
\end{table*}

\paragraph{\textbf{Discussion}}

\texttt{AERIAL} is competitive to state-of-the-art baselines in original SMAC,  indicating that replacing the true state $s_{t}$ with the joint memory representation $\mathbf{h_{t}}$ does not notably harm performance.  Despite outperforming most baselines in some maps,  we do not claim significant outperformance here,  since we regard most SMAC maps as widely solved by the community anyway \cite{ellissmacv2}.

\subsection{MessySMAC}\label{subsec:messy_smac_results}

\paragraph{\textbf{Setting}}
We evaluate \texttt{AERIAL} in MessySMAC using the same maps as in Section \ref{subsec:smac_results}.  We set $\phi = 15\%$ and $K = 10$.

\begin{figure*}
\centering
     \subfigure[\texttt{3s5z}\label{fig:3s5z_results}]{%
         \includegraphics[width=0.32\textwidth]{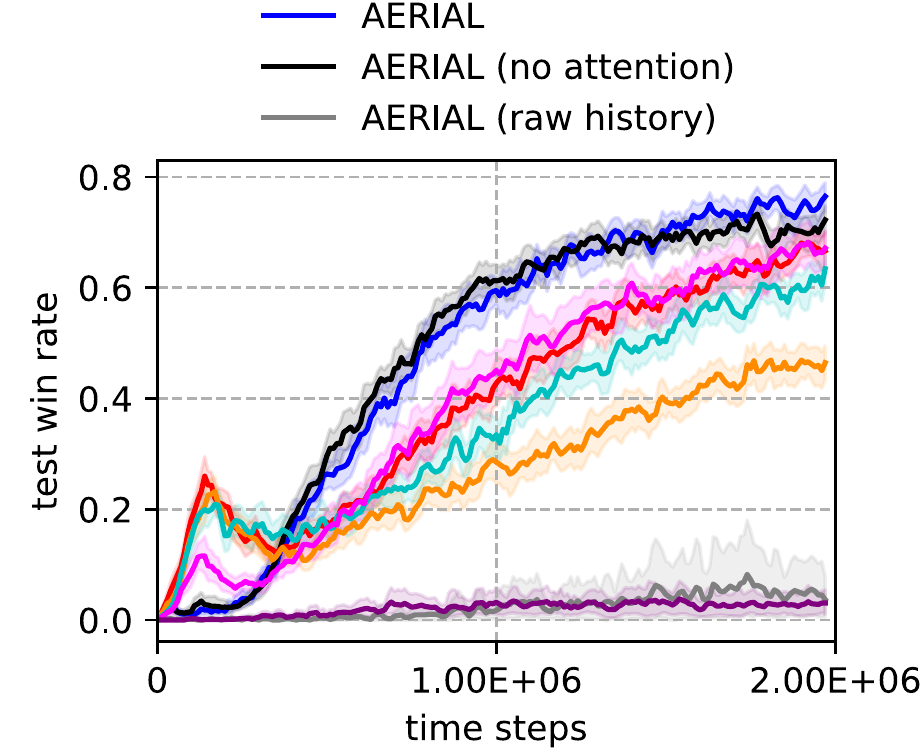}
     }\hfill
     \subfigure[\texttt{10m\_vs\_11m}\label{fig:10m_vs_11m_results}]{%
         \includegraphics[width=0.32\textwidth]{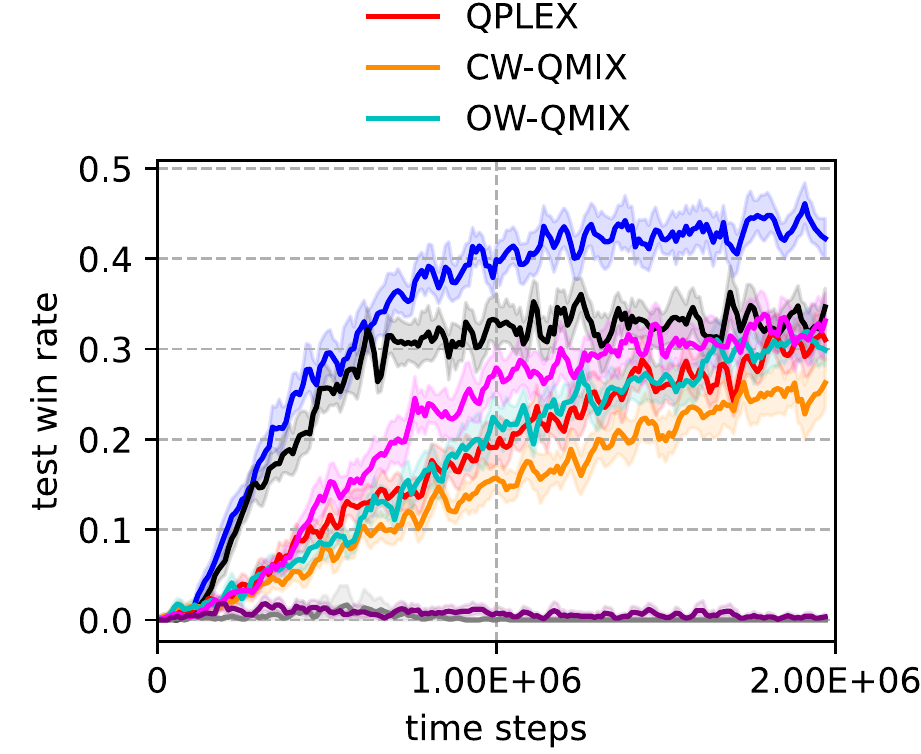}
     }\hfill
     \subfigure[\texttt{2c\_vs\_64zg}\label{fig:2c_vs_64zg_results}]{%
         \includegraphics[width=0.32\textwidth]{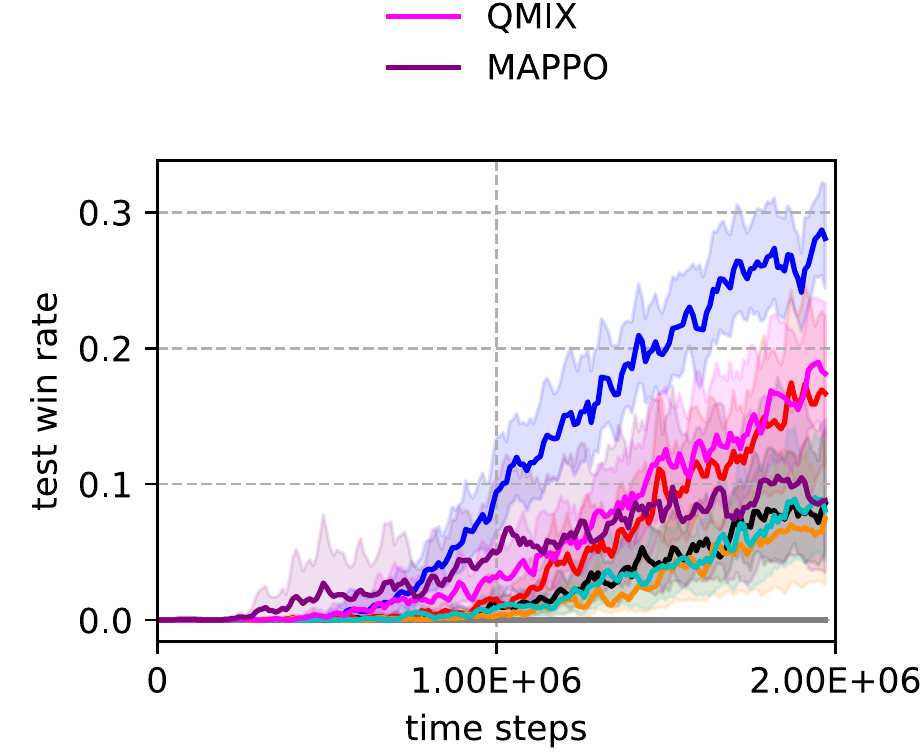}
     }\\
     \subfigure[\texttt{3s\_vs\_5z}\label{fig:3s_vs_5z_results}]{%
         \includegraphics[width=0.32\textwidth]{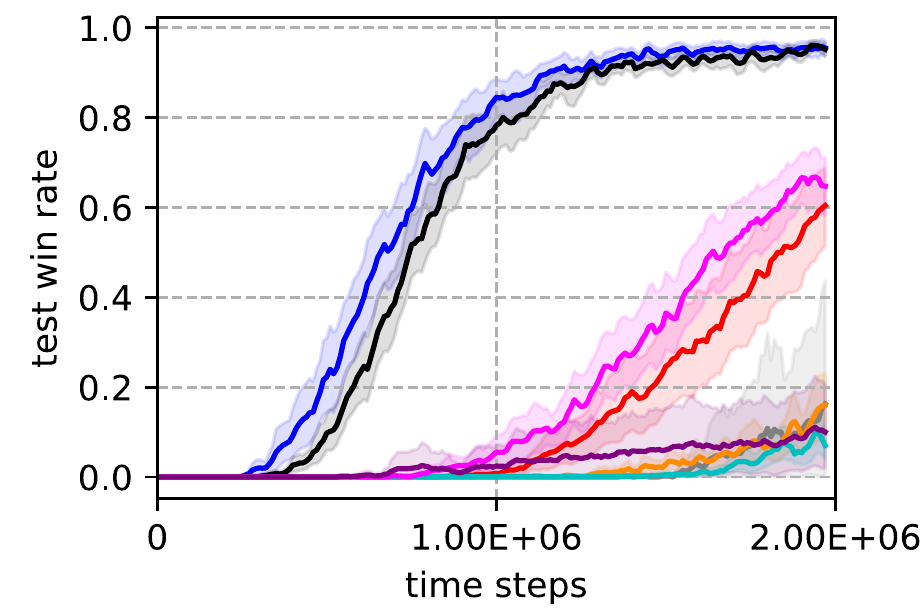}
     }\hfill
     \subfigure[\texttt{5m\_vs\_6m}\label{fig:5m_vs_6m_results}]{%
         \includegraphics[width=0.32\textwidth]{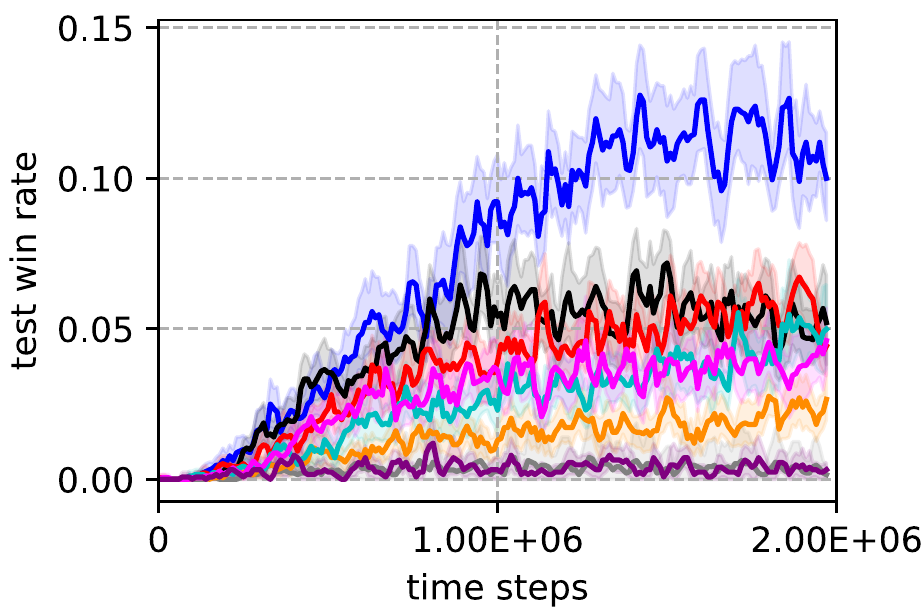}
     }\hfill
     \subfigure[\texttt{3s5z\_vs\_3s6z}\label{fig:3s5z_vs_3s6z_results}]{%
         \includegraphics[width=0.32\textwidth]{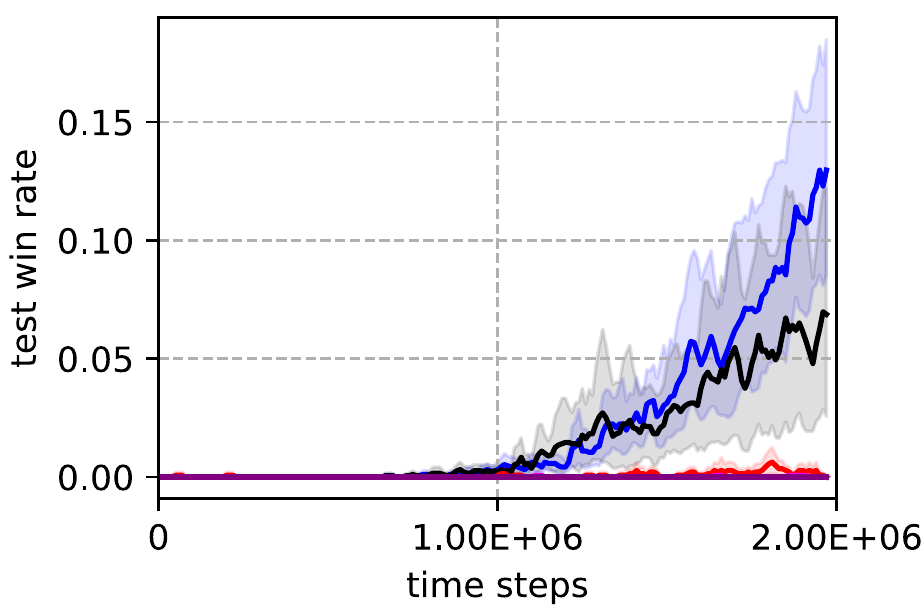}
     }
     \caption{Average learning progress w.r.t.  the win rate of \texttt{AERIAL} variants and state-of-the-art baselines in MessySMAC for 2 million steps over 20 runs.  Shaded areas show the 95\% confidence interval.  The legend at the top applies across all plots.}
     \label{fig:messy_sc2_results}
\end{figure*}

\paragraph{\textbf{Results}}
The results are shown in Fig.  \ref{fig:messy_sc2_results}.  \texttt{AERIAL} performs best in all maps with \texttt{AERIAL (no attention)} being second best except in \texttt{2c\_vs\_64zg}.  In \texttt{3s5z\_vs\_3s6z},  only \texttt{AERIAL} and \texttt{AERIAL (no attention)} progress notably.  \texttt{AERIAL (raw history)} performs worst in all maps.  \texttt{MAPPO} only progresses notably in \texttt{2c\_vs\_64zg}.

\paragraph{\textbf{Discussion}}

Similar to the Dec-Tiger experiment,  the results confirm the benefit of exploiting more accurate closed-loop information in domains with stochastic partial observability.  \texttt{AERIAL} consistently outperforms \texttt{AERIAL (no attention)},  indicating that self-attention can correct for the naive independence assumption of all $h_{t,i} \in \mathbf{h_{t}}$.  \texttt{MAPPO} performs especially poorly in MessySMAC due to its misleading dependence on true states without any credit assignment,  confirming the findings of \cite{ellissmacv2}.

\subsection{Robustness against Stochastic Partial Observability}\label{subsec:results_state_uncertainty}

\paragraph{\textbf{Setting}}
To evaluate the robustness of \texttt{AERIAL} and \texttt{AERIAL (no attention)} against various stochasticity configurations in MessySMAC,  we manipulate $\Omega$ through the observation negation probability $\phi$ and $b_{0}$ through the number of initial random steps $K$ as defined in Section \ref{subsec:messy_smac}.  We compare the results with \texttt{QMIX} and \texttt{QPLEX} as the best performing state-of-the-art baselines in MessySMAC according to the results in Section \ref{subsec:messy_smac_results}.  We present summarized plots,  where the results are aggregated accross all maps from Section \ref{subsec:messy_smac_results}. 
To avoid that easy maps dominate the average win rate,  since all approaches achieve high values there,  we normalize the values by the maximum win rate achieved in the respective map for all tested configurations of $\phi$ and $K$.  Thus,  we ensure an equal weighting regardless of the particular difficulty level.  If not mentioned otherwise,  we set $\phi = 15\%$ and $K = 10$ as default parameters based on Section \ref{subsec:messy_smac_results}.

\begin{figure}
\centering
     \subfigure[normalized test win rate\label{fig:win_rate_observations}]{%
         \includegraphics[width=0.23\textwidth]{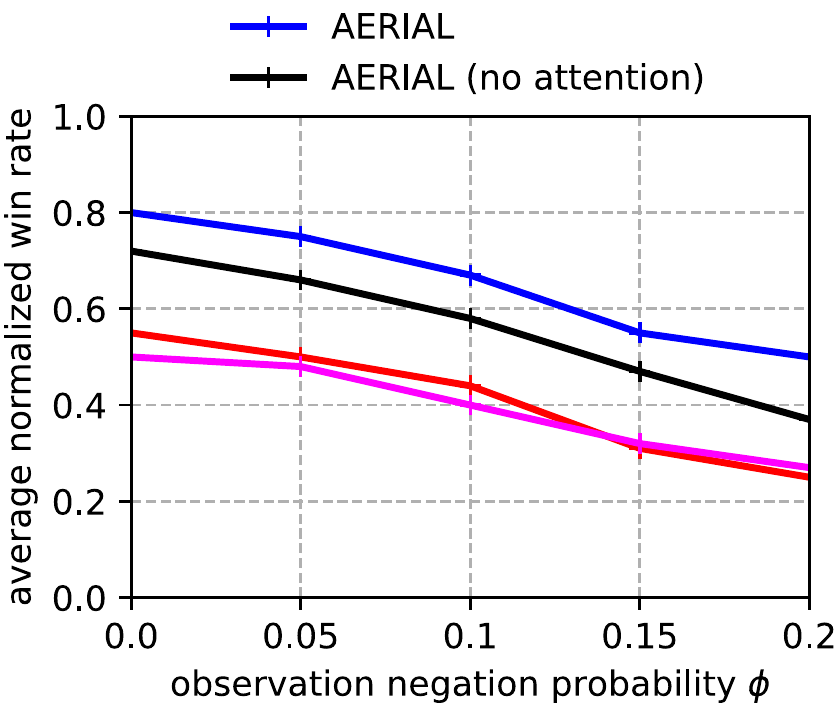}
     }\hfill
     \subfigure[\# maps best out of 6\label{fig:best_observations}]{%
         \includegraphics[width=0.22\textwidth]{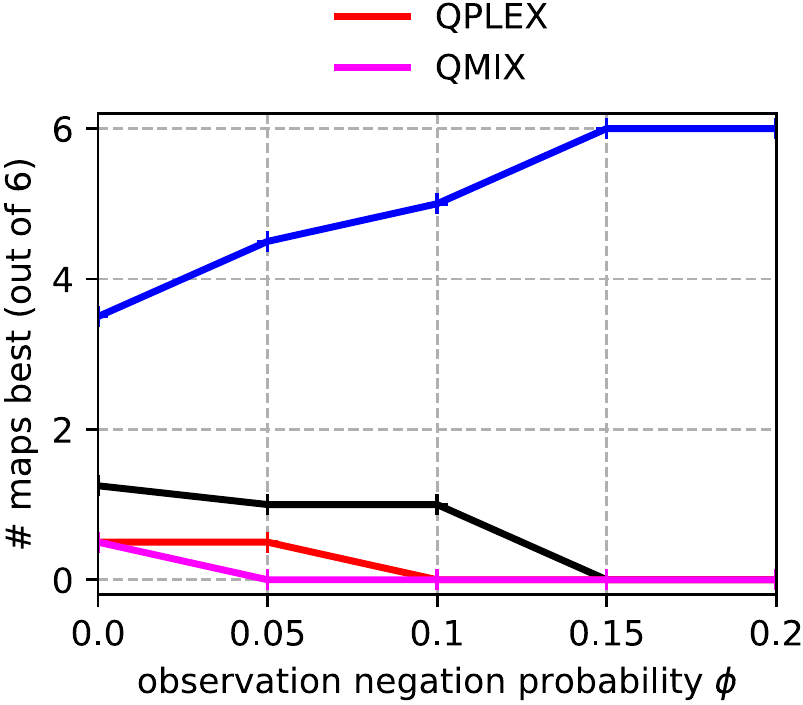}
     }
     \caption{Evaluation of \texttt{AERIAL},  \texttt{AERIAL (no attention)},  and the best MessySMAC baselines for different observation negation probabilities $\phi$ affecting observation stochasticity w.r.t.  $\Omega$ (20 runs per configuration).  (a) The average normalized test win rate accross all 6 MessySMAC maps from Section \ref{subsec:messy_smac_results}.  (b) The number of maps best out of 6.  The legend at the top applies across all plots.}
     \label{fig:state_uncertainty_observation_failures}
\end{figure}

\paragraph{\textbf{Results}}
The results regarding observation stochasticity w.r.t.  $\Omega$ and $\phi$ are shown in Fig.  \ref{fig:state_uncertainty_observation_failures}.  Fig.  \ref{fig:win_rate_observations} shows that the average win rates of all approaches decrease with increasing $\phi$ with \texttt{AERIAL} consistently achieving the highest average win rate in all configurations.  Fig.  \ref{fig:best_observations} shows that \texttt{AERIAL} performs best in most MessySMAC maps,  especially when $\phi \geq 15\%$.  \texttt{AERIAL (no attention)} performs second best.

\begin{figure}
\centering
     \subfigure[normalized test win rate\label{fig:win_rate_states}]{%
         \includegraphics[width=0.23\textwidth]{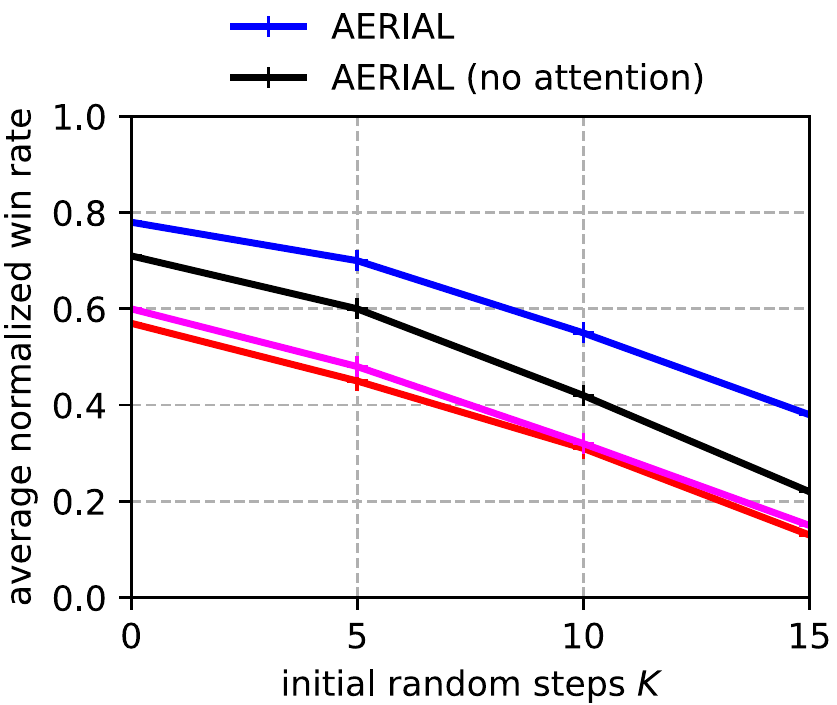}
     }\hfill
     \subfigure[\# maps best out of 6\label{fig:best_states}]{%
         \includegraphics[width=0.22\textwidth]{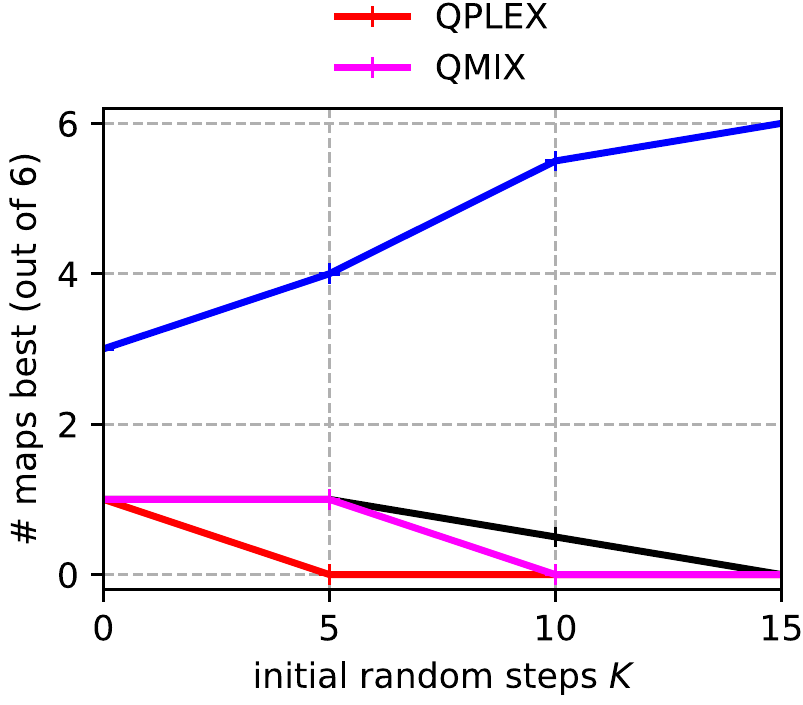}
     }
     \caption{Evaluation of \texttt{AERIAL},  \texttt{AERIAL (no attention)},  and the best MessySMAC baselines for different initial random steps $K$ affecting initialization stochasticity w.r.t.  $b_{0}$ (20 runs per configuration).  (a) The average normalized test win rate accross all 6 MessySMAC maps from Section \ref{subsec:messy_smac_results}.  (b) The number of maps best out of 6.  The legend at the top applies across all plots.}
     \label{fig:state_uncertainty_initial_states}
\end{figure}

The results regarding initialization stochasticity w.r.t.  $b_{0}$ and $K$ are shown in Fig.   \ref{fig:state_uncertainty_initial_states}.  Analogously to Fig.  \ref{fig:state_uncertainty_observation_failures},  Fig.  \ref{fig:win_rate_states} shows that the average (normalized) win rates of all approaches decrease with increasing $K$ with \texttt{AERIAL} consistently achieving the highest average win rate in all configurations.  Fig.  \ref{fig:best_states} shows that \texttt{AERIAL} performs best in most MessySMAC maps,  especially when $K \geq 10$.  \texttt{AERIAL (no attention)} performs second best.

\paragraph{\textbf{Discussion}}

Our results systematically demonstrate the robustness of \texttt{AERIAL} and \texttt{AERIAL (no attention)} against various stochasticity configurations according to $\Omega$ and $b_{0}$.  State-based CTDE is notably less effective in settings,  where observation and initialization stochasticity is high.  As \texttt{AERIAL} consistently performs best in all maps when $\phi \geq 15\%$ or $K \geq 10$,  we conclude that providing an adequate representation of $\mathbf{P}^{\boldsymbol\pi}(\boldsymbol\tau_{\mathbf{t}}|b_{0})$ according to Eq. \ref{eq:history_probability} that is learned,  e.g.,  through $\mathbf{h_{t}}$ and self-attention,  is more beneficial for CTDE than merely relying on true states when facing domains with high stochastic partial observability.

\section{Conclusion and Future Work}

To tackle general multi-agent problems, which are messy and only observable through noisy sensors, we need adequate algorithms and benchmarks that sufficiently consider stochastic partial observability.

In this paper, we proposed AERIAL to approximate value functions under stochastic partial observability with a learned representation of multi-agent recurrence, considering more accurate closed-loop information about decentralized agent decisions than state-based CTDE.

We then introduced \emph{MessySMAC},  a modified version of SMAC with stochastic observations and higher variance in initial states,  to provide a more general and configurable Dec-POMDP benchmark regarding stochastic partial observability.  We showed visually in Fig.  \ref{fig:initial_state_visualization} and experimentally in Section \ref{sec:experiments} that MessySMAC scenarios pose a greater challenge than their original SMAC counterparts due to observation and initialization stochasticity.

Compared to state-based CTDE,  AERIAL offers a simple but effective approach to general Dec-POMDPs,  being competitive in original SMAC and superior in Dec-Tiger and MessySMAC,  which both exhibit observation and initialization stochasticity unlike original SMAC.  Simply replacing the true state with memory representations can already improve performance in most scenarios,  confirming the need for more accurate closed-loop information about decentralized agent decisions.  Self-attention can correct for the naive independence assumption of agent-wise recurrence to further improve performance,  especially when observation or initialization stochasticity is high.

We plan to further evaluate AERIAL in SMACv2 and mixed competitive-cooperative settings with multiple CTDE instances \cite{lowe2017multi,phan2020learning}.  

\section*{Acknowledgements}

This work was partially funded by the Bavarian Ministry for Economic Affairs, Regional Development and Energy as part of a project to support the thematic development of the Institute for Cognitive Systems.

\bibliography{references}

\begin{thebibliography}{39}
\providecommand{\natexlab}[1]{#1}
\providecommand{\url}[1]{\texttt{#1}}
\expandafter\ifx\csname urlstyle\endcsname\relax
  \providecommand{\doi}[1]{doi: #1}\else
  \providecommand{\doi}{doi: \begingroup \urlstyle{rm}\Url}\fi

\bibitem[Amato et~al.(2007)Amato, Bernstein, and
  Zilberstein]{amato2007optimizing}
Amato, C., Bernstein, D.~S., and Zilberstein, S.
\newblock {Optimizing} {Memory}-{Bounded} {Controllers} for {Decentralized}
  {POMDPs}.
\newblock In \emph{Proceedings of the 23rd Conference on Uncertainty in
  Artificial Intelligence}, pp.\  1--8, 2007.

\bibitem[Bernstein et~al.(2005)Bernstein, Hansen, and
  Zilberstein]{bernstein2005bounded}
Bernstein, D.~S., Hansen, E.~A., and Zilberstein, S.
\newblock {Bounded} {Policy} {Iteration} for {Decentralized} {POMDPs}.
\newblock In \emph{IJCAI}, pp.\  52--57, 2005.

\bibitem[Boutilier(1996)]{boutilier1996planning}
Boutilier, C.
\newblock {Planning}, {Learning} and {Coordination} in {Multiagent} {Decision}
  {Processes}.
\newblock In \emph{Proceedings of the 6th conference on Theoretical aspects of
  rationality and knowledge}, pp.\  195--210. Morgan Kaufmann Publishers Inc.,
  1996.

\bibitem[Chen et~al.(2021)Chen, Lu, Rajeswaran, Lee, Grover, Laskin, Abbeel,
  Srinivas, and Mordatch]{chen2021decision}
Chen, L., Lu, K., Rajeswaran, A., Lee, K., Grover, A., Laskin, M., Abbeel, P.,
  Srinivas, A., and Mordatch, I.
\newblock {Decision} {Transformer}: {Reinforcement} {Learning} via {Sequence}
  {Modeling}.
\newblock In Ranzato, M., Beygelzimer, A., Dauphin, Y., Liang, P., and Vaughan,
  J.~W. (eds.), \emph{Advances in Neural Information Processing Systems},
  volume~34, pp.\  15084--15097. Curran Associates, Inc., 2021.

\bibitem[Cho et~al.(2014)Cho, van Merri{\"e}nboer, Bahdanau, and
  Bengio]{cho2014properties}
Cho, K., van Merri{\"e}nboer, B., Bahdanau, D., and Bengio, Y.
\newblock {On} the {Properties} of {Neural} {Machine} {Translation}:
  {Encoder}-{Decoder} {Approaches}.
\newblock In \emph{Proceedings of SSST-8, Eighth Workshop on Syntax, Semantics
  and Structure in Statistical Translation}, pp.\  103--111, 2014.

\bibitem[Ellis et~al.(2022)Ellis, Moalla, Samvelyan, Sun, Mahajan, Foerster,
  and Whiteson]{ellissmacv2}
Ellis, B., Moalla, S., Samvelyan, M., Sun, M., Mahajan, A., Foerster, J.~N.,
  and Whiteson, S.
\newblock {SMACv2}: {An} {Improved} {Benchmark} for {Cooperative}
  {Multi}-{Agent} {Reinforcement} {Learning}.
\newblock 2022.

\bibitem[Emery-Montemerlo et~al.(2004)Emery-Montemerlo, Gordon, Schneider, and
  Thrun]{emery2004approximate}
Emery-Montemerlo, R., Gordon, G., Schneider, J., and Thrun, S.
\newblock {Approximate} {Solutions} for {Partially} {Observable} {Stochastic}
  {Games} with {Common} {Payoffs}.
\newblock In \emph{Proceedings of the Third International Joint Conference on
  Autonomous Agents and Multiagent Systems - Volume 1}, AAMAS '04, pp.\
  136–143, USA, 2004. IEEE Computer Society.
\newblock ISBN 1581138644.

\bibitem[Foerster et~al.(2018)Foerster, Farquhar, Afouras, Nardelli, and
  Whiteson]{foerster2017counterfactual}
Foerster, J., Farquhar, G., Afouras, T., Nardelli, N., and Whiteson, S.
\newblock {Counterfactual} {Multi}-{Agent} {Policy} {Gradients}.
\newblock \emph{Proceedings of the AAAI Conference on Artificial Intelligence},
  32\penalty0 (1), Apr. 2018.

\bibitem[Gupta et~al.(2017)Gupta, Egorov, and
  Kochenderfer]{gupta2017cooperative}
Gupta, J.~K., Egorov, M., and Kochenderfer, M.
\newblock {Cooperative} {Multi}-{Agent} {Control} using {Deep} {Reinforcement}
  {Learning}.
\newblock In \emph{Autonomous Agents and Multiagent Systems}, pp.\  66--83.
  Springer, 2017.

\bibitem[Hausknecht \& Stone(2015)Hausknecht and Stone]{hausknecht2015deep}
Hausknecht, M. and Stone, P.
\newblock {Deep} {Recurrent} {Q}-{Learning} for {Partially} {Observable}
  {MDPs}.
\newblock In \emph{2015 AAAI Fall Symposium Series}, 2015.

\bibitem[Hochreiter \& Schmidhuber(1997)Hochreiter and
  Schmidhuber]{hochreiter1997long}
Hochreiter, S. and Schmidhuber, J.
\newblock {Long} {Short}-{Term} {Memory}.
\newblock \emph{Neural Computation}, 9\penalty0 (8):\penalty0 1735--1780, 1997.

\bibitem[Hu \& Foerster(2019)Hu and Foerster]{hu2019simplified}
Hu, H. and Foerster, J.~N.
\newblock {Simplified} {Action} {Decoder} for {Deep} {Multi}-{Agent}
  {Reinforcement} {Learning}.
\newblock In \emph{International Conference on Learning Representations}, 2019.

\bibitem[Iqbal \& Sha(2019)Iqbal and Sha]{iqbal19a}
Iqbal, S. and Sha, F.
\newblock {Actor}-{Attention}-{Critic} for {Multi}-{Agent} {Reinforcement}
  {Learning}.
\newblock In Chaudhuri, K. and Salakhutdinov, R. (eds.), \emph{Proceedings of
  the 36th International Conference on Machine Learning}, volume~97 of
  \emph{Proceedings of Machine Learning Research}, pp.\  2961--2970, Long
  Beach, California, USA, 09--15 Jun 2019. PMLR.

\bibitem[Iqbal et~al.(2021)Iqbal, De~Witt, Peng, Boehmer, Whiteson, and
  Sha]{iqbal2021randomized}
Iqbal, S., De~Witt, C. A.~S., Peng, B., Boehmer, W., Whiteson, S., and Sha, F.
\newblock {Randomized} {Entity}-wise {Factorization} for {Multi}-{Agent}
  {Reinforcement} {Learning}.
\newblock In Meila, M. and Zhang, T. (eds.), \emph{Proceedings of the 38th
  International Conference on Machine Learning}, volume 139 of
  \emph{Proceedings of Machine Learning Research}, pp.\  4596--4606. PMLR,
  18--24 Jul 2021.

\bibitem[Kaelbling et~al.(1998)Kaelbling, Littman, and
  Cassandra]{kaelbling1998planning}
Kaelbling, L.~P., Littman, M.~L., and Cassandra, A.~R.
\newblock {Planning} and {Acting} in {Partially} {Observable} {Stochastic}
  {Domains}.
\newblock \emph{Artificial intelligence}, 101\penalty0 (1-2):\penalty0 99--134,
  1998.

\bibitem[Khan et~al.(2022)Khan, Ahmed, and Sukthankar]{khan2022transformer}
Khan, M.~J., Ahmed, S.~H., and Sukthankar, G.
\newblock {Transformer}-{Based} {Value} {Function} {Decomposition} for
  {Cooperative} {Multi}-{Agent} {Reinforcement} {Learning} in {StarCraft}.
\newblock \emph{Proceedings of the AAAI Conference on Artificial Intelligence
  and Interactive Digital Entertainment}, 18\penalty0 (1):\penalty0 113--119,
  Oct. 2022.

\bibitem[Lowe et~al.(2017)Lowe, Wu, Tamar, Harb, Abbeel, and
  Mordatch]{lowe2017multi}
Lowe, R., Wu, Y., Tamar, A., Harb, J., Abbeel, P., and Mordatch, I.
\newblock {Multi}-{Agent} {Actor}-{Critic} for {Mixed}
  {Cooperative}-{Competitive} {Environments}.
\newblock In Guyon, I., Luxburg, U.~V., Bengio, S., Wallach, H., Fergus, R.,
  Vishwanathan, S., and Garnett, R. (eds.), \emph{Advances in Neural
  Information Processing Systems}, volume~30. Curran Associates, Inc., 2017.

\bibitem[Lyu et~al.(2021)Lyu, Xiao, Daley, and Amato]{lyu2021contrasting}
Lyu, X., Xiao, Y., Daley, B., and Amato, C.
\newblock {Contrasting} {Centralized} and {Decentralized} {Critics} in
  {Multi}-{Agent} {Reinforcement} {Learning}.
\newblock In \emph{Proceedings of the 20th International Conference on
  Autonomous Agents and Multiagent Systems}, pp.\  844--852, 2021.

\bibitem[Lyu et~al.(2022)Lyu, Baisero, Xiao, and Amato]{lyu2022deeper}
Lyu, X., Baisero, A., Xiao, Y., and Amato, C.
\newblock {A} {Deeper} {Understanding} of {State}-{Based} {Critics} in
  {Multi}-{Agent} {Reinforcement} {Learning}.
\newblock \emph{Proceedings of the AAAI Conference on Artificial Intelligence},
  36\penalty0 (9):\penalty0 9396--9404, Jun. 2022.
\newblock \doi{10.1609/aaai.v36i9.21171}.

\bibitem[Nair et~al.(2003)Nair, Tambe, Yokoo, Pynadath, and
  Marsella]{nair2003taming}
Nair, R., Tambe, M., Yokoo, M., Pynadath, D., and Marsella, S.
\newblock {Taming} {Decentralized} {POMDPs}: {Towards} {Efficient} {Policy}
  {Computation} for {Multiagent} {Settings}.
\newblock In \emph{Proceedings of the 18th International Joint Conference on
  Artificial Intelligence}, IJCAI'03, pp.\  705–711, San Francisco, CA, USA,
  2003. Morgan Kaufmann Publishers Inc.

\bibitem[Oliehoek \& Amato(2016)Oliehoek and Amato]{oliehoek2016concise}
Oliehoek, F.~A. and Amato, C.
\newblock \emph{{A} {Concise} {Introduction} to {Decentralized} {POMDPs}},
  volume~1.
\newblock Springer, 2016.

\bibitem[Oliehoek et~al.(2008)Oliehoek, Spaan, and
  Vlassis]{oliehoek2008optimal}
Oliehoek, F.~A., Spaan, M.~T., and Vlassis, N.
\newblock {Optimal} and {Approximate} {Q}-{Value} {Functions} for
  {Decentralized} {POMDPs}.
\newblock \emph{Journal of Artificial Intelligence Research}, 32:\penalty0
  289--353, 2008.

\bibitem[Phan et~al.(2020)Phan, Gabor, Sedlmeier, Ritz, Kempter, Klein, Sauer,
  Schmid, Wieghardt, Zeller, et~al.]{phan2020learning}
Phan, T., Gabor, T., Sedlmeier, A., Ritz, F., Kempter, B., Klein, C., Sauer,
  H., Schmid, R., Wieghardt, J., Zeller, M., et~al.
\newblock {Learning} and {Testing} {Resilience} in {Cooperative}
  {Multi}-{Agent} {Systems}.
\newblock In \emph{Proceedings of the 19th International Conference on
  Autonomous Agents and Multiagent Systems}, AAMAS '20, pp.\  1055–1063.
  International Foundation for Autonomous Agents and Multiagent Systems, 2020.

\bibitem[Phan et~al.(2021)Phan, Ritz, Belzner, Altmann, Gabor, and
  Linnhoff-Popien]{phan2021vast}
Phan, T., Ritz, F., Belzner, L., Altmann, P., Gabor, T., and Linnhoff-Popien,
  C.
\newblock {VAST}: {Value} {Function} {Factorization} with {Variable} {Agent}
  {Sub}-{Teams}.
\newblock In Ranzato, M., Beygelzimer, A., Dauphin, Y., Liang, P., and Vaughan,
  J.~W. (eds.), \emph{Advances in Neural Information Processing Systems},
  volume~34, pp.\  24018--24032. Curran Associates, Inc., 2021.

\bibitem[Phan et~al.(2023)Phan, Ritz, N\"{u}\ss{}lein, K\"{o}lle, Gabor, and
  Linnhoff-Popien]{phan2022attention2}
Phan, T., Ritz, F., N\"{u}\ss{}lein, J., K\"{o}lle, M., Gabor, T., and
  Linnhoff-Popien, C.
\newblock {Attention}-{Based} {Recurrency} for {Multi}-{Agent} {Reinforcement}
  {Learning} under {State} {Uncertainty}.
\newblock In \emph{Extended Abstracts of the 22nd International Conference on
  Autonomous Agents and Multiagent Systems}, AAMAS '23, pp.\  2839–2841.
  International Foundation for Autonomous Agents and Multiagent Systems, 2023.
\newblock ISBN 9781450394321.

\bibitem[Rashid et~al.(2018)Rashid, Samvelyan, Schroeder, Farquhar, Foerster,
  and Whiteson]{rashid2018qmix}
Rashid, T., Samvelyan, M., Schroeder, C., Farquhar, G., Foerster, J., and
  Whiteson, S.
\newblock {QMIX}: {Monotonic} {Value} {Function} {Factorisation} for {Deep}
  {Multi}-{Agent} {Reinforcement} {Learning}.
\newblock In Dy, J. and Krause, A. (eds.), \emph{Proceedings of the 35th
  International Conference on Machine Learning}, volume~80 of \emph{Proceedings
  of Machine Learning Research}, pp.\  4295--4304. PMLR, 10--15 Jul 2018.

\bibitem[Rashid et~al.(2020)Rashid, Farquhar, Peng, and
  Whiteson]{rashid2020weighted}
Rashid, T., Farquhar, G., Peng, B., and Whiteson, S.
\newblock {Weighted} {QMIX}: {Expanding} {Monotonic} {Value} {Function}
  {Factorisation} for {Deep} {Multi}-{Agent} {Reinforcement} {Learning}.
\newblock In Larochelle, H., Ranzato, M., Hadsell, R., Balcan, M.~F., and Lin,
  H. (eds.), \emph{Advances in Neural Information Processing Systems},
  volume~33, pp.\  10199--10210. Curran Associates, Inc., 2020.

\bibitem[Samvelyan et~al.(2019)Samvelyan, Rashid, Schroeder~de Witt, Farquhar,
  Nardelli, Rudner, Hung, Torr, Foerster, and Whiteson]{samvelyan2019starcraft}
Samvelyan, M., Rashid, T., Schroeder~de Witt, C., Farquhar, G., Nardelli, N.,
  Rudner, T.~G., Hung, C.-M., Torr, P.~H., Foerster, J., and Whiteson, S.
\newblock {The} {StarCraft} {Multi}-{Agent} {Challenge}.
\newblock In \emph{Proceedings of the 18th International Conference on
  Autonomous Agents and Multiagent Systems}, AAMAS '19, pp.\  2186–2188,
  Richland, SC, 2019. International Foundation for Autonomous Agents and
  Multiagent Systems.
\newblock ISBN 9781450363099.

\bibitem[Son et~al.(2019)Son, Kim, Kang, Hostallero, and Yi]{son2019qtran}
Son, K., Kim, D., Kang, W.~J., Hostallero, D.~E., and Yi, Y.
\newblock {QTRAN}: {Learning} to {Factorize} with {Transformation} for
  {Cooperative} {Multi}-{Agent} {Reinforcement} {Learning}.
\newblock In Chaudhuri, K. and Salakhutdinov, R. (eds.), \emph{Proceedings of
  the 36th International Conference on Machine Learning}, volume~97 of
  \emph{Proceedings of Machine Learning Research}, pp.\  5887--5896. PMLR,
  09--15 Jun 2019.

\bibitem[Sunehag et~al.(2018)Sunehag, Lever, Gruslys, Czarnecki, Zambaldi,
  Jaderberg, Lanctot, Sonnerat, Leibo, Tuyls, and Graepel]{sunehag2017value}
Sunehag, P., Lever, G., Gruslys, A., Czarnecki, W.~M., Zambaldi, V., Jaderberg,
  M., Lanctot, M., Sonnerat, N., Leibo, J.~Z., Tuyls, K., and Graepel, T.
\newblock {Value}-{Decomposition} {Networks} for {Cooperative} {Multi}-{Agent}
  {Learning} based on {Team} {Reward}.
\newblock In \emph{Proceedings of the 17th International Conference on
  Autonomous Agents and Multiagent Systems}, AAMAS '18, pp.\  2085–2087,
  Richland, SC, 2018. International Foundation for Autonomous Agents and
  Multiagent Systems.

\bibitem[Szer et~al.(2005)Szer, Charpillet, and Zilberstein]{szer2005maa}
Szer, D., Charpillet, F., and Zilberstein, S.
\newblock {MAA}*: {A} {Heuristic} {Search} {Algorithm} for {Solving}
  {Decentralized} {POMDPs}.
\newblock UAI'05, pp.\  576–583, Arlington, Virginia, USA, 2005. AUAI Press.
\newblock ISBN 0974903914.

\bibitem[Tan(1993)]{tan1993multi}
Tan, M.
\newblock {Multi}-{Agent} {Reinforcement} {Learning}: {Independent} versus
  {Cooperative} {Agents}.
\newblock In \emph{Proceedings of the Tenth International Conference on
  International Conference on Machine Learning}, ICML'93, pp.\  330–337, San
  Francisco, CA, USA, 1993. Morgan Kaufmann Publishers Inc.
\newblock ISBN 1558603077.

\bibitem[Vaswani et~al.(2017)Vaswani, Shazeer, Parmar, Uszkoreit, Jones, Gomez,
  Kaiser, and Polosukhin]{vaswani2017attention}
Vaswani, A., Shazeer, N., Parmar, N., Uszkoreit, J., Jones, L., Gomez, A.~N.,
  Kaiser, L.~u., and Polosukhin, I.
\newblock {Attention} is {All} {You} {Need}.
\newblock In Guyon, I., Luxburg, U.~V., Bengio, S., Wallach, H., Fergus, R.,
  Vishwanathan, S., and Garnett, R. (eds.), \emph{Advances in Neural
  Information Processing Systems}, volume~30. Curran Associates, Inc., 2017.

\bibitem[Vinyals et~al.(2019)Vinyals, Babuschkin, Czarnecki, Mathieu, Dudzik,
  Chung, Choi, Powell, Ewalds, Georgiev, et~al.]{vinyals2019grandmaster}
Vinyals, O., Babuschkin, I., Czarnecki, W.~M., Mathieu, M., Dudzik, A., Chung,
  J., Choi, D.~H., Powell, R., Ewalds, T., Georgiev, P., et~al.
\newblock {Grandmaster} {Level} in {StarCraft} {II} using {Multi}-{Agent}
  {Reinforcement} {Learning}.
\newblock \emph{Nature}, pp.\  1--5, 2019.

\bibitem[Wang et~al.(2021)Wang, Ren, Liu, Yu, and Zhang]{wang2020qplex}
Wang, J., Ren, Z., Liu, T., Yu, Y., and Zhang, C.
\newblock {QPLEX}: {Duplex} {Dueling} {Multi}-{Agent} {Q}-{Learning}.
\newblock In \emph{International Conference on Learning Representations}, 2021.

\bibitem[Watkins \& Dayan(1992)Watkins and Dayan]{watkins1992q}
Watkins, C.~J. and Dayan, P.
\newblock {Q}-{Learning}.
\newblock \emph{Machine Learning}, 8\penalty0 (3-4):\penalty0 279--292, 1992.

\bibitem[Wen et~al.(2022)Wen, Kuba, Lin, Zhang, Wen, Wang, and
  Yang]{wen2022multi}
Wen, M., Kuba, J.~G., Lin, R., Zhang, W., Wen, Y., Wang, J., and Yang, Y.
\newblock {Multi}-{Agent} {Reinforcement} {Learning} is a {Sequence} {Modeling}
  {Problem}.
\newblock \emph{arXiv preprint arXiv:2205.14953}, 2022.

\bibitem[Whittlestone et~al.(2021)Whittlestone, Arulkumaran, and
  Crosby]{whittlestone2021societal}
Whittlestone, J., Arulkumaran, K., and Crosby, M.
\newblock {The} {Societal} {Implications} of {Deep} {Reinforcement} {Learning}.
\newblock \emph{Journal of Artificial Intelligence Research}, 70:\penalty0
  1003–1030, May 2021.
\newblock ISSN 1076-9757.
\newblock \doi{10.1613/jair.1.12360}.

\bibitem[Yu et~al.(2022)Yu, Velu, Vinitsky, Gao, Wang, Bayen, and
  Wu]{yu2022the}
Yu, C., Velu, A., Vinitsky, E., Gao, J., Wang, Y., Bayen, A., and Wu, Y.
\newblock {The} {Surprising} {Effectiveness} of {PPO} in {Cooperative}
  {Multi}-{Agent} {Games}.
\newblock In \emph{36th Conference on Neural Information Processing Systems
  Datasets and Benchmarks Track}, 2022.

\end{thebibliography}
\bibliographystyle{icml2023}

\newpage
\appendix
\onecolumn
\section{Limitations and Societal Impacts}

\subsection{Limitations}

AERIAL does not significantly outperform state-of-the-art baselines in easier domains without stochastic partial observability as indicated by the original SMAC results in Table \ref{tab:smac_results},  implying that simplified Dec-POMDP settings might benefit from more specialized algorithms.
The dependence on joint memory representations $\mathbf{h_{t}} =  \langle h_{t,i} \rangle_{i \in \mathcal{D}}$ might induce some bias w.r.t.  agent behavior policies which could limit performance in hard exploration domains therefore requiring additional mechanisms beyond the scope of this work.  The full version of AERIAL requires additional compute\footnote{The additional amount regarding wall clock time was negligible in our experiments though.} due to the transformer component in Fig.  \ref{fig:aerial_architecture} which can be compensated by using a more (parameter) efficient value function factorization operator $\Psi$,  e.g.,  using QMIX instead of QPLEX.

\subsection{Potential Negative Societal Impacts}
The goal of our work is to realize autonomous systems to solve complex tasks under stochastic partial observability as motivated in Section \ref{sec:introduction}.  We refer to \cite{whittlestone2021societal} for a general overview regarding societal implications of deep RL and completely focus on cooperative MARL settings in the following.

AERIAL is based on a centralized training regime to learn decentralized policies with a common objective.  That objective might include bias of a central authority and could potentially harm opposing parties, e.g., via discrimination or misleading information.  Since training is conducted in a laboratory or a simulation,  the resulting system might exhibit unsafe or questionable behavior when being deployed in the real world due to poor generalization, e.g.,  leading to accidents or unfair decisions. The transformer component in Fig.  \ref{fig:aerial_architecture} might require a significant amount of additional compute for tuning and training therefore increasing overall cost.  The self-attention weights of Eq.  \ref{eq:self_attention} could be used to discriminate participating individuals in an unethical way,  e.g.,  discarding less relevant groups of individuals according to the softmax output.

Similar to original SMAC,  MessySMAC is based on team battles,  indicating that any MARL algorithm mastering that challenge could be misused for real combat,  e.g.,  in autonomous weapon systems to realize distributed and coordinated strategies.  Since MessySMAC covers the aspect of stochastic partial observability,  successfully evaluated algorithms could be potentially more effective and dangerous in real-world scenarios.

\section{Dec-Tiger Example}\label{subsec:appendix_dec_tiger}

Given the Dec-Tiger example from Section \ref{subsec:state_vs_history} with a horizon of $T=2$,  the tiger being behind the right door ($s_{R}$),  and both agents having listened in the first step,  where agent $1$ heard $z_{L}$ and agent $2$ heard $z_{R}$: The final state-based values are defined by $Q_{\textit{MDP}}^{*}(s_{t},\mathbf{a_t}) = \mathcal{R}(s_{t}, \mathbf{a_t})$.

Due to both agents perceiving different observations,  i.e.,  $z_{L}$ and $z_{R}$ respectively,  the probability of being in state $s_{R}$ is 50\% according to the belief state, i.e.,  $b(s_{R}|\boldsymbol\tau_{\mathbf{t}}) = b(s_{L}|\boldsymbol\tau_{\mathbf{t}}) = \frac{1}{2}$.  Thus,  the true optimal Dec-POMDP values for the final time step are defined by:
\begin{align}
\begin{split}
Q^{*}(\boldsymbol\tau_{\mathbf{t}},  \mathbf{a_t}) &=  \sum_{s_{t} \in \mathcal{S}} b(s_{t}|\boldsymbol\tau_{\mathbf{t}})\mathcal{R}(s_{t}, \mathbf{a_t})\\
						&= \frac{1}{2}(Q_{\textit{MDP}}^{*}(s_{L},  \mathbf{a_t}) + Q_{\textit{MDP}}^{*}(s_{R},  \mathbf{a_t}))
\end{split}
\end{align}

The values of $Q_{\textit{MDP}}^{*}$ and $Q^{*}$ for the final time step $t = 2$ in the example are given in Table \ref{tab:dec_tiger_values}.  Both agents can reduce the expected penalty when always performing the same action.  Therefore,  it is likely for MARL to converge to a joint policy that recommends the same actions for both agents,  especially when synchronization techniques like parameter sharing are used \cite{tan1993multi,gupta2017cooperative,yu2022the}.

\begin{table}
\centering
\caption{The values of $Q_{\textit{MDP}}^{*}$ and $Q^{*}$ for the final time step $t = 2$ in the Dec-Tiger example from Section \ref{subsec:state_vs_history}.}
\begin{tabular}{|c||c|c|c|} \hline
$\mathbf{a_t}$ & $Q_{\textit{MDP}}^{*}(s_{L},  \mathbf{a_t})$ & $Q_{\textit{MDP}}^{*}(s_{R},  \mathbf{a_t})$ & $Q^{*}(\boldsymbol\tau_{\mathbf{t}},  \mathbf{a_t})$\\ \hline
$\color{blue}\mathbf{\langle \textit{li},  \textit{li} \rangle}$ & $\color{blue}\mathbf{-2}$ & $\color{blue}\mathbf{-2}$ & $\color{blue}\mathbf{-2}$\\
$\langle \textit{li},  o_L \rangle$ & -101 & +9 & -46\\
$\langle \textit{li},  o_R \rangle$ & +9 & -101 & -46\\
$\langle o_L,  \textit{li} \rangle$ & -101 & +9 & -46\\
$\color{red}\mathbf{\langle o_L,  o_L \rangle}$ & $\color{red}\mathbf{-50}$ & $\color{red}\mathbf{+20}$ & $\color{red}\mathbf{-15}$\\
$\langle o_L,  o_R \rangle$ & -100 & -100 & -100\\
$\langle o_R,  \textit{li} \rangle$ & +9 & -101 & -46\\
$\langle o_R,  o_L \rangle$ & -100 & -100 & -100\\
$\color{red}\mathbf{\langle o_R,  o_R \rangle}$ & $\color{red}\mathbf{+20}$ & $\color{red}\mathbf{-50}$ & $\color{red}\mathbf{-15}$\\\hline
\end{tabular}\label{tab:dec_tiger_values}
\end{table}

\section{Full Algorithm of AERIAL}\label{sec:appendix_aerial_full_algorithm}

The complete formulation of AERIAL is given in Algorithm \ref{algorithm:AERIAL}.  Note that AERIAL does not depend on true states $s_t$ at all,  since the experience samples $e_{t}$ (Line 23) used for training do not record any states.

\begin{algorithm}
\caption{Attention-based Embeddings of Recurrence In multi-Agent Learning (AERIAL)}\label{algorithm:AERIAL}
\begin{algorithmic}[1]
\STATE Initialize parameters for $\langle Q_{i} \rangle_{i \in \mathcal{D}}$ and $\Psi$.
\FOR{episode $m \leftarrow 1,E$}
\STATE Sample $s_{0}$, $\mathbf{z_{0}}$, and $\boldsymbol\tau_{\mathbf{0}}$ via $b_{0}$ and $\Omega$
\FOR{time step $t \leftarrow 0,T-1$}
\FOR{agent $i \in \mathcal{D}$}
	\STATE $a_{t,i} \leftarrow \pi_{i}(\tau_{t,i})$ \algorithmiccomment{Use $\textit{argmax}_{a_{t,i}\in\mathcal{A}_{i}} Q_{i}(\tau_{t,i}, a_{t,i})$}
	\STATE $\textit{rand} \sim U(0,1)$\algorithmiccomment{Sample from uniform distribution}
	\IF{$\textit{rand} \leq \epsilon$}
		\STATE Select random action $a_{t,i} \in \mathcal{A}_{i}$ \algorithmiccomment{Explore with $\epsilon$-greedy}
	\ENDIF
\ENDFOR
\STATE $\mathbf{a_{t}} \leftarrow \langle a_{t,i} \rangle_{i \in \mathcal{D}}$
\STATE Execute joint action $\mathbf{a_{t}}$
\STATE $s_{t+1} \sim \mathcal{T}(s_{t+1}|s_{t}, \mathbf{a_{t}})$
\STATE $\mathbf{z_{t+1}} \sim \Omega(\mathbf{z_{t+1}}|\mathbf{a_{t}},s_{t+1})$
\STATE $\mathbf{h_{t}} \leftarrow  \langle h_{t,i} \rangle_{i \in \mathcal{D}}$ \algorithmiccomment{Query memory representations of all agents}
\STATE Detach $\mathbf{h_{t}}$ from computation graph\algorithmiccomment{Avoid additional differentiation through $\Psi$ or Eq. \ref{eq:self_attention}}
\STATE $\boldsymbol\tau_{\mathbf{t+1}} \leftarrow \langle \boldsymbol\tau_{\mathbf{t}},  \mathbf{a_{t}},  \mathbf{z_{t+1}} \rangle$ \algorithmiccomment{Concatenate $\boldsymbol\tau_{\mathbf{t}}$,  $\mathbf{a_{t}}$,  and $\mathbf{z_{t+1}}$}
	\FOR{attention head $c \leftarrow 1,C$}
			\STATE $\textit{attention}_{c} \leftarrow \textit{att}_{c}(\mathbf{h_{t}})$ \algorithmiccomment{Process individual recurrences according to Eq. \ref{eq:self_attention}}
		\ENDFOR
\STATE $\textit{rec}_{t} \leftarrow \textit{MLP}(\sum_{c=1}^{C} \textit{attention}_{c})$ \algorithmiccomment{See Section \ref{subsec:aerial_concept}}
\STATE $e_{t} \leftarrow \langle \boldsymbol\tau_{\mathbf{t}},  \mathbf{a_{t}},  r_{t},  \mathbf{z_{t+1}},  \textit{rec}_{t} \rangle$
\STATE Store experience sample $e_{t}$
\ENDFOR
\STATE Train $\Psi$ and $\langle Q_{i} \rangle_{i \in \mathcal{D}}$ using all $e_{t}$ \algorithmiccomment{See Fig.  \ref{fig:aerial_architecture}}
\ENDFOR
\end{algorithmic}
\end{algorithm}

\newpage
\section{Experiment Details}\label{subsec:appendix_technical}

\subsection{Computing infrastructure}

All training and test runs were performed in parallel on a computing cluster of fifteen x86\_64 GNU/Linux (Ubuntu 18.04.5 LTS) machines with i7-8700 @ 3.2GHz CPU (8 cores) and 64 GB RAM.  We did not use any GPU in our experiments.

\subsection{Hyperparameters and Neural Network Architectures}\label{subsec:hyperparameters}

Our experiments are based on PyMARL and the code from \cite{rashid2020weighted} under the Apache License 2.0.  We use the default setting from the paper without further hyperparameter tuning as well as the same neural network architectures for the agent RNNs,  i.e.,  \emph{gated recurrent units (GRU)} of \cite{cho2014properties} with 64 units,  and the respective factorization operators $\Psi$ as specified by default for each state-of-the-art baseline in Section \ref{sec:experiments}.  We set the loss weight $\alpha = 0.75$ for \texttt{CW-QMIX} and \texttt{OW-QMIX}. 

For \texttt{MAPPO},  we use the hyperparameters suggested in \cite{yu2022the} for SMAC, where we set the clipping parameter to 0.1 and use an epoch count of 5.   The parameter $\lambda$ for generalized advantage estimation is set to 1.  The centralized critic has two hidden layers of 128 units with ReLU activation,  a single linear output,  and conditions on \emph{agent-specific global states} which concatenate the global state and the individual observation per agent.  The policy network of MAPPO has a similar recurrent architecture like the local utility functions $Q_i$ and additionally applies softmax to the output layer.

\texttt{AERIAL} is implemented using \texttt{QMIX} as factorization operator $\Psi$ according to Fig.  \ref{fig:aerial_architecture}.  We also experimented with \texttt{QPLEX} as alternative with no significant difference in performance.  Thus,  we stick with \texttt{QMIX} for computational efficiency due to fewer trainable parameters.  The transformer has $C=4$ heads $c \in \{1,  ...,  C\}$ with respective MLPs $W_{q}^{c}$,  $W_{k}^{c}$,  and $W_{v}^{c}$,  each having one hidden layer of $d_{\textit{att}} = 64$ units with ReLU activation.  The three subsequent MLP layers of Line 22 in Algorithm \ref{algorithm:AERIAL} have 64 units with ReLU activation.

All neural networks are trained using RMSProp with a learning rate of 0.0005.

\end{document}